\documentclass[fleqn,usenatbib]{mnras}

\usepackage{newtxtext,newtxmath}
\usepackage{multirow}
\usepackage{multicol}
\usepackage{natbib}  

\usepackage[T1]{fontenc}

\DeclareRobustCommand{\VAN}[3]{#2}
\let\VANthebibliography\thebibliography
\def\thebibliography{\DeclareRobustCommand{\VAN}[3]{##3}\VANthebibliography}

\usepackage{graphicx}	
\usepackage{amsmath}	

\title[]{The Nature of Classical Be Star Outbursts: A Multi-Epoch Study of the Be system EPIC 202060631}

\author[GaoQing et al.]{
Gao Qing,$^{1}$
Yuan HaiLong$^{2}$
\\
$^{1}$Independent Researcher\\
$^{2}$Key Laboratory of Optical Astronomy, National Astronomical Observatories,  Chinese Academy of Sciences, 20A Datun Road, \\ Beijing, 100101, People’s Republic of China
}

\date{}

\pubyear{}

\begin{document}
\label{firstpage}
\pagerange{\pageref{firstpage}--\pageref{lastpage}}
\maketitle

\begin{abstract}
Through cross-matching the ASAS-SN photometric light curves and LAMOST spectroscopic
observations, we serendipitously captured a rare major outburst event from the Be star EPIC
202060631, lasting over 1000 days. Fortuitously, the LAMOST low-resolution spectra densely covered
the flux rising stage with 20 epochs, while an additional 7 low-resolution and 11 medium-resolution 
spectra monitored the subsequent decay phase.
Moreover, this target was observed by \emph{Kepler} telescope in its K2 mission when before
the outburst and by \emph{TESS} telescope while after returning to quiescence.
Analyses of these datasets reveal striking phenomena accompanying the outburst,
including a $\sim$30\% brightness increase in optical bands,
dramatic changes in pulsation behavior and mode amplitudes, indications of radial and tangential motions in the photosphere, and clear evidence of mass ejection and circumstellar disk formation. We modeled the central star using the BRUCE04 code and analyzed the disk structure using HDUST. Our results suggest episodic mass injection events from the central star triggered the disk buildup and carried imprints of the changing stellar pulsations. This study offers unique insights into the connections between photospheric activities, disk evolution, and stellar rotation during Be star outbursts.
\end{abstract}

\begin{keywords}
Classical Be Star -- Light Curve -- Spectra
\end{keywords}



\section{Introduction}

Be stars are rapidly rotating B-type stars that exhibit Balmer emission lines arising from circumstellar decretion disks 
(Rivinius et al. 2013).
They display a wide range of variability, including non-radial pulsations and irregular outbursts thought to be related to disk buildup and dissipation cycles. During outburst events lasting from weeks to years, the disk density and size change substantially, leading to significant photometric brightening and spectroscopic variations. Studying the detailed photometric and spectroscopic evolution of Be stars throughout an entire outburst cycle provides crucial clues for unraveling the physical mechanisms driving the mass outflow dynamics and disk formation.

However, continuous spectroscopic monitoring spanning complete outburst episodes has traditionally been challenging due to the serendipitous nature of these events. The advent of wide-field time-domain photometric surveys has enabled discoveries of rare outbursts, but adequate spectroscopic follow-up remains lacking. In this work, we present a remarkable case where extensive time-series photometry and spectroscopy fortuitously captured all stages of a prolonged outburst from the Be star EPIC 202060631.

By probing changes in pulsation modes, projected rotational velocities, radial velocities, emission line strengths, and disk density profiles, we aim to shed new light on the intricate connections between photospheric activities, circumstellar disk evolution, and the central star's rotation during strong outburst episodes of Be stars. Our multi-epoch observations across a wide wavelength range, coupled with detailed modeling, offer a unique opportunity to study the mass ejection process and disk formation mechanisms in rapidly rotating early-type stars.

\section{Observations}

\begin{figure*}
\includegraphics[width=1.0\linewidth]{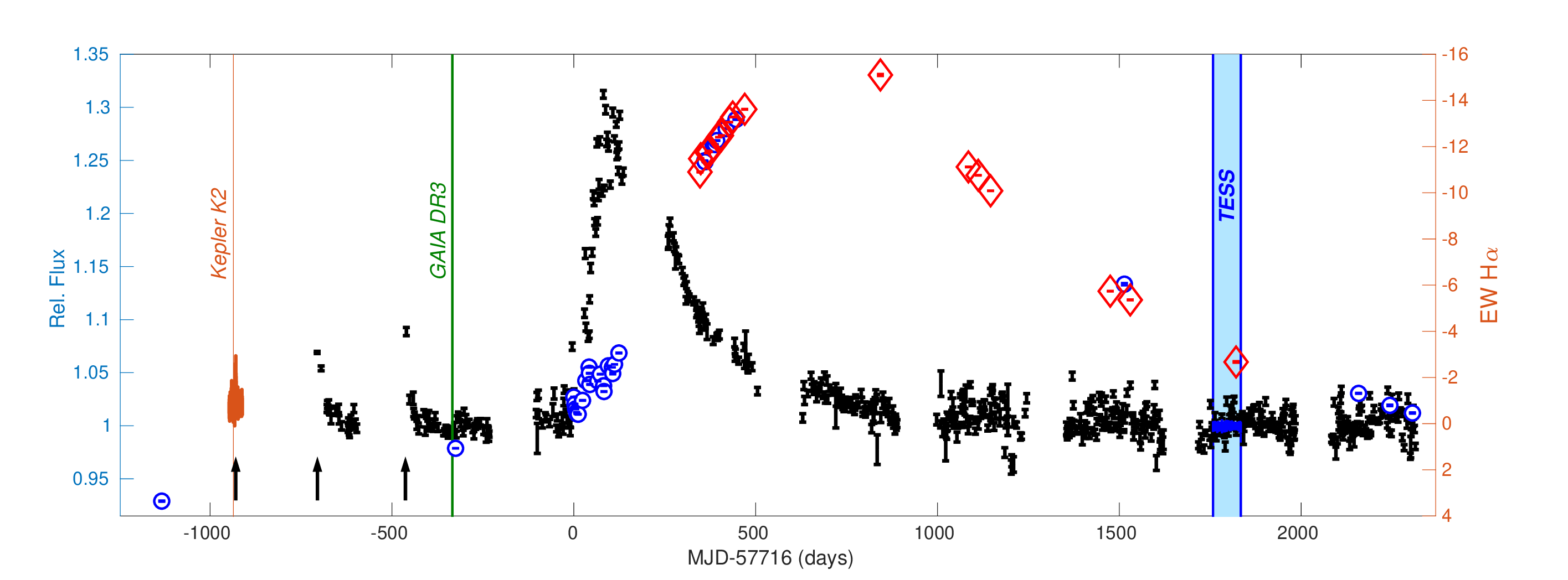}
\caption{\label{Allobs} All observations of the target EPIC 202060631 with time relative to MJD 57716.
Left axis: Kepler K2 light curve (red), ASAS-SN photometric data (black error bars), TESS light curve (blue curve and highlighted as light blue region).
Three minor outbursts are marked with black arrows.
Right axis: Equivalent Widths (EWs) of H$\alpha$ emission lines calculated from LAMOST spectra.
Values from low-resolution spectra are marked as blue open circles, 
while those from medium resolution spectra are red diamonds.
Errors are smaller than the marker size, appearing as short dashes.
}
\end{figure*}

\begin{table*}
\centering
\caption{\label{TabObs} The Observation Log of LAMOST Spectroscopic Data for the Target at Both Low and Medium Resolutions.
Columns 1-5 list the observed Modified Julian Date (MJD), observation ID number, corresponding burst period relative to photometric observations, spectrum filename, and the mean signal-to-noise ratio (SNR) across the wavelength range, respectively. 
The mean SNR provides a general indication of the spectrum quality.}
\begin{tabular}{ccccc}
\hline
MJD & obsID & Period & SpectraName & meanSNR \\
\hline
\multicolumn{5}{c}{\textit{Low Resolution}} \\
\hline
56583 & 1 & QS & spec-56583-GAC093N22V2\_sp04-043 & 450 \\
57391 & 2 & -- & spec-57391-KP061029N225952V01\_sp04-224 & 102 \\
57716 & 3 & LR & spec-57716-HIP29425K201\_sp04-009 & 923 \\
57719 & 4 & LR & spec-57719-HIP29425K201\_sp04-009 & 1002 \\
57721 & 5 & LR & spec-57721-HIP29425K201\_sp04-009 & 1155 \\
57724 & 6 & LR & spec-57724-HIP29425K201\_sp04-009 & 1148 \\
57725 & 7 & LR & spec-57725-HIP29425K201\_sp04-009 & 1294 \\
57728 & 8 & LR & spec-57728-HIP29425K201\_sp04-009 & 992 \\
57740 & 9 & LR & spec-57740-HIP29425K201\_sp04-009 & 1107 \\
57749 & 10 & LR & spec-57749-HIP29425K201\_sp04-009 & 740 \\
57758 & 11 & LR & spec-57758-HIP29425K201\_sp04-009 & 773 \\
57759 & 12 & LR & spec-57759-HIP29425K201\_sp04-009 & 897 \\
57760 & 13 & LR & spec-57760-HIP29425K201\_sp04-009 & 950 \\
57782 & 14 & LR & spec-57782-HIP29425K201\_sp04-009 & 664 \\
57789 & 15 & LR & spec-57789-HIP29425K201\_sp04-009 & 1203 \\
57798 & 16 & LR & spec-57798-HIP29425K201\_sp04-009 & 570 \\
57799 & 17 & LR & spec-57799-HIP29425K201\_sp04-009 & 851 \\
57811 & 18 & LR & spec-57811-HIP29425K201\_sp04-009 & 1214 \\
57822 & 19 & LR & spec-57822-HIP29425K201\_sp04-009 & 775 \\
57823 & 20 & LR & spec-57823-HIP29425K201\_sp04-009 & 376 \\
57828 & 21 & LR & spec-57828-HIP29425K201\_sp04-009 & 641 \\
57840 & 22 & LR & spec-57840-HIP29425K201\_sp04-009 & 780 \\
58076 & 23 & LD & spec-58076-HIP29425K2\_sp04-009 & 926 \\
58077 & 24 & LD & spec-58077-HIP29425K2\_sp04-009 & 705 \\
58099 & 25 & LD & spec-58099-HIP29425K201\_sp04-009 & 876 \\
58105 & 26 & LD & spec-58105-HIP29425K201\_sp04-009 & 1081 \\
58109 & 27 & LD & spec-58109-HIP29425K201\_sp04-009 & 1069 \\
58135 & 28 & LD & spec-58135-HIP29425K201\_sp04-009 & 993 \\
58162 & 29 & LD & spec-58162-HIP29425K201\_sp04-009 & 831 \\
58788 & 30 & new-QS & spec-58788-HIP29425K201\_sp04-009 & 906 \\
59230 & 31 & new-QS & spec-59230-TD061156N231225K01\_sp04-009 & 276 \\
59874 & 32 & new-QS & spec-59874-TD061156N231225K01\_sp04-151 & 641 \\
59960 & 33 & new-QS & spec-59960-TD061156N231225K01\_sp04-151 & 386 \\
60022 & 34 & new-QS & spec-60022-TD061156N231225K01\_sp04-151 & 521 \\
\hline
\multicolumn{5}{c}{\textit{Medium Resolution}} \\
\hline
58063 & 1 & LD & med-58063-HIP29425K201\_sp04-009 & 129 \\
58065 & 2 & LD & med-58065-HIP29425K201\_sp04-009 & 164 \\
58084 & 3 & LD & med-58084-HIP29425K201\_sp04-009 & 194 \\
58086 & 4 & LD & med-58086-HIP29425K201\_sp04-009 & 275 \\
58088 & 5 & LD & med-58088-HIP29425K201\_sp04-009 & 249 \\
58114 & 6 & LD & med-58114-HIP29425K201\_sp04-009 & 260 \\
58122 & 7 & LD & med-58122-HIP29425K201\_sp04-009 & 192 \\
58144 & 8 & LD & med-58144-HIP29425K201\_sp04-009 & 237 \\
58153 & 9 & LD & med-58153-HIP29425K201\_sp04-009 & 221 \\
58186 & 10 & LD & med-58186-HIP29425K201\_sp04-009 & 123 \\
58559 & 11 & LD & med-58559-TD060648N233818B01\_sp08-037 & 60 \\
58801 & 12 & new-QS & med-58801-TD060648N233818B01\_sp08-037 & 188 \\
58828 & 13 & new-QS & med-58828-HIP30046H342501\_sp10-014 & 313 \\
58862 & 14 & new-QS & med-58862-TD060648N233818B01\_sp08-037 & 223 \\
59191 & 15 & new-QS & med-59191-TD061156N231225K01\_sp04-009 & 145 \\
59246 & 16 & new-QS & med-59246-TD061156N231225K01\_sp04-009 & 99 \\
59537 & 17 & new-QS & med-59537-TD060648N233818B01\_sp08-037 & 51 \\
\hline
\end{tabular}
\end{table*}

Regarding the Observation and Data Reduction section, I have summarized it as follows: The target has a Kepler K2 ID of EPIC202060631, located at RA= 92.82, DEC=23.42, and V mag of 10.75. It exhibited a major outburst (MAB) in the V-band photometers from the All-Sky Automated Survey for SuperNovae (ASAS-SN) datasets \footnote{https://asas-sn.osu.edu/photometry}. The MAB began around the Modified Julian Day (MJD) 57716 and lasted more than 1000 days, with up to a 30 percent flux increase at its maximum. Subsequently, the target fell back to a new quiescent (new-QS) state until now.

All observations collected for EPIC 202060631 are shown in Fig.~\ref{Allobs}. 
The photometric data consists of three components: ASAS-SN, Kepler K2 \footnote{https://archive.stsci.edu/k2/epic/search.php}, and TESS \footnote{https://mast.stsci.edu/tesscut/}. The ASAS-SN commenced on 2014 December 16, with photometry acquired in the V and g bands. The offset between these two bands was corrected by minimizing the linear dispersion within the overlapping V and g observations over 100 days. Multiple exposures within one day were binned to increase the signal-to-noise ratio (SNR). The white-light photometry of Kepler K2 started 239 days before the ASAS-SN observation and lasted for 36 days. As shown in the light blue region in Fig.~\ref{Allobs}, the Be star was also observed by the TESS telescope with ID TIC45726822 while the target had fallen back to quiescence. The three consecutive sectors, namely 43, 44, and 45, were combined into a single dataset with a total baseline of 76 days.

As pointed out by black arrows in Fig.~\ref{Allobs}, with the combination of Kepler K2 and ASAS-SN, three minor outbursts are concentrated around -940, -706, and -460 days relative to MJD 57716. The light brightened above quiescent levels by about 5 to 10 percent, with a duration of several days. Moreover, the minor bursts repeated approximately every 240 days, and a 4th minor burst may have occurred but was not captured during the ASAS-SN observational gaps. This semi-periodicity is a primary characteristic of Be star outbursts, such as the Be bursts observed in targets like KIC 9715425, KIC 6954726, and $\Omega$ CMa.

Serendipitously, the LAMOST low-resolution observations (R$\sim$1500) captured one quiescent (QS) spectrum before all outbursts and 20 observations during the light rising (LR) period of the MAB within 124 days, 7 observations during the light decay (LD) period within 86 days, and 5 spectra taken when the target reached its new quiescent (new-QS) state in optical fluxes. Observation details are shown in Table.~\ref{TabObs}, and its equivalent widths (EWs) variations of H$\alpha$ are shown as blue circles in Fig.~\ref{Allobs}. It should be noted that the 4th spectrum from the bottom did not have the H$\alpha$ region released, so no EW is shown in Fig.~\ref{Allobs} for that observation.

Furthermore, a total of 17 medium-resolution (R$\sim$7200) spectra were observed with relatively high SNRs. Details are shown in Table.~\ref{TabObs}. The first 10 spectra overlap the observation time of the low-resolution spectra, which were taken during the middle of the LD period. The observation at MJD 58559 reached the maximum H$\alpha$ emission level, corresponding to the end of the LD phase. The remaining spectra are all from the new-QS period, and their EWs are shown as red diamonds in Fig.~\ref{Allobs}.

\section{Modeling the Central star with BRUCE04}

With the first low-resolution and emission line-free LAMOST spectrum taken during the quiescent state (QS) before the outburst, we can preliminarily explore the pure stellar parameters by fitting the non-LTE Tlusty BSTAR2006 template with the UlySS package. 
This reveals an effective temperature of $\sim$18000 K, log g $\sim$3.8, and a projected rotational velocity ($v$sin\,$i$) of $\sim$200 km\,s$^{-1}$.

\begin{figure}
\includegraphics[width=1.0\linewidth]{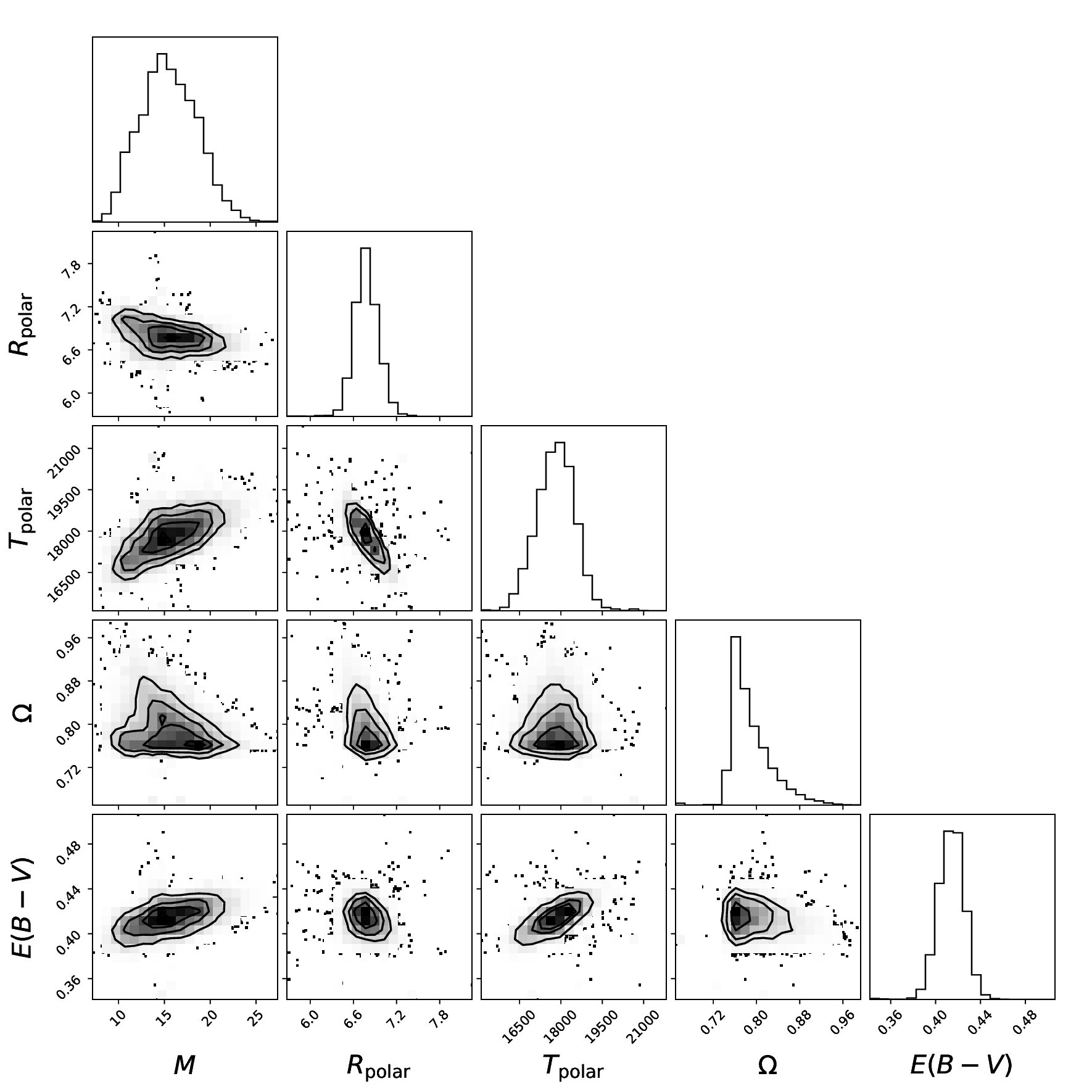}
\caption{\label{br4mcmc} Posterior probability distributions from the MCMC sampling results of modeling EPIC 202060631 with the BRUCE04 model.
The independent physical parameters are stellar mass, polar radius, polar temperature, rotation rate ($\Omega$), and reddening E(B-V) to match the observed SED from Gaia photometry. The inclination angle was fixed at 45 degrees.
}
\end{figure}

\begin{table}
\centering
  \begin{tabular}{l r  l r}
  \hline
  \hline
  \noalign{\smallskip}
   &BRUCE04 Model & & \\
  \noalign{\smallskip}
  \hline
  \noalign{\smallskip}
  $M$ ($M_{\odot}$) & $15.3^{+3.3}_{-3.0}$ & $R_{\rm polar}$ ($R_{\odot}$) & $6.8^{+0.2}_{-0.1}$ \\
  \noalign{\smallskip}
  $L$ ($L_{\odot}$) & $3632^{+456}_{-420}$ & $i\,(^{\rm o})$ & \textcolor{red}{45}  \\
  \noalign{\smallskip}
  $T_{\rm eff}$ (K) & $16526^{+644}_{-664}$ & $T_{\rm eq}$ (K) &  $14827^{+699}_{-714}$ \\
  \noalign{\smallskip}
  $T_{\rm polar}$ (K) & $17833^{+634}_{-706}$ & $V_{\rm eq}$ (km/s) & $326^{+28}_{-44}$ \\
  \noalign{\smallskip}
  log\,$g_{\rm polar}$ (cgs) & $3.96^{+0.09}_{-0.11}$ & log\,$g_{\rm eq}$ (cgs) &  $3.60^{+0.12}_{-0.14}$ \\
  \noalign{\smallskip}
   ${R_{\rm eq}} / {R_{\rm polar}}$ & $1.13^{+0.03}_{-0.01}$ & $\Omega / \Omega_{\rm c}$ & $0.78^{+0.05}_{-0.02}$ \\
   \noalign{\smallskip}
   $\rm P_{\rm rot}$ (days) & $1.2^{+0.2}_{-0.1}$ &  E(B-V) &  $0.41^{+0.01}_{-0.01}$\\
   \hline
  \end{tabular}
   \caption{The median values and 68.3\% confidence intervals (14th and 86th percentiles) for the physical parameters obtained from MCMC sampling by fitting the first quiescent period spectra and Gaia SED of EPIC 202060631. The inclination angle was fixed at 45 degrees.}
    \label{tabBr4}
\end{table}

Additionally, the target has Gaia EDR3 data released, with an observation ID of 3425411417005784448. The broadband fluxes are summarized as follows:

\begin{itemize}

\item Gbp band (5040 \AA): ($1.82 \pm 0.12) \times 10^{-13}$ erg~s$^{-1}$~cm$^{-2}$~\AA$^{-1}$

\item G band (5820 \AA): $1.38 \times 10^{-13}$ erg~s$^{-1}$~cm$^{-2}$~\AA$^{-1}$

\item Grp band (7620 \AA): ($7.95 \pm 0.52) \times 10^{-13}$ erg~s$^{-1}$~cm$^{-2}$~\AA$^{-1}$

\end{itemize}

In addition to the first LAMOST spectrum, the Gaia DR3 photometry also indicates the star's spectral energy distribution (SED) during the QS period, 
which is clearly shown in Fig.~\ref{Allobs}. 
Furthermore, the geometric distance provided by the 
Bailer-Jones et al. (2021) catalog is 2575$^{+133}_{-195}$ pc. 
Therefore, stellar parameters such as mass, radius, and luminosity can be deduced.

However, traditional spherical stellar atmosphere models are known to be inaccurate for Be stars due to their rapid rotation and ellipsoidal shapes. 
The BRUCE04 model incorporates stellar oblateness when constructing synthetic spectra of early-type stars. 
For the BRUCE04 model of a rigidly rotating Be star, 
four independent parameters need to be specified: 
mass (M), polar radius (R), polar temperature (T$_{\rm p}$), 
and rotational velocity ($\Omega/\Omega_{\rm c}$).
We also included the system inclination angle, interstellar extinction E(B-V), 
and relative radial velocity as free parameters. 
We ran Markov Chain Monte Carlo (MCMC) simulations with emcee to find the best-fitting parameters by 
matching the model spectra to the quiescent LAMOST spectrum and Gaia G$_{\rm bp}$ and G$_{\rm rp}$ photometry.

From the morphology of the H$\alpha$ emission line, empirically, the system should be at a median inclination angle. 
Therefore, we decided to fix the inclination angle at 45 degrees. 
Utilizing the BRUCE04 model with the fixed inclination angle and employing the MCMC from emcee, 
the MCMC converged nicely, as shown in Fig.~\ref{br4mcmc}. The specific parameters and distributions are presented in Table.~\ref{tabBr4}. 
The derived E(B-V) value is consistent with estimates from the 3D dust maps given by 
Green et al. (2019), with a value of E(g-r) = 0.44 $\pm$ 0.02.

\section{Pulsation behavior changes before and after burst observed by Kepler K2 and TESS}

\begin{figure}
\includegraphics[width=1.0\linewidth]{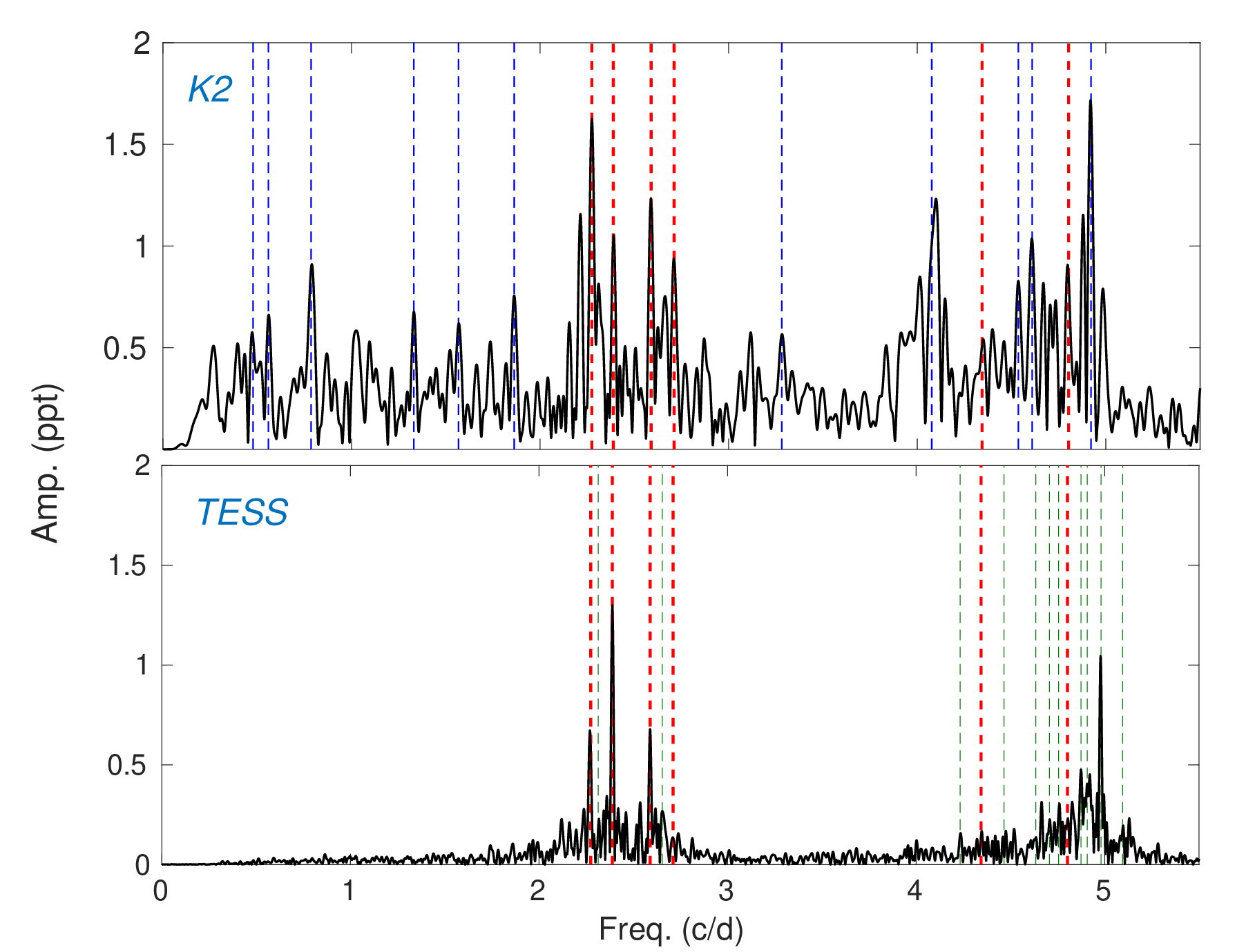}
\caption{\label{PowSpec} Amplitude spectra in parts per thousand (ppt) derived from the Kepler K2 light curve (top panel) and the TESS light curve (bottom panel). 
The six common signal peaks are marked with red dashed lines. 
Signals detected only in K2 are marked with blue dashed lines, 
while those found only in TESS are marked with green dashed lines. 
The specific signal parameters are listed in Table.~\ref{tabMod}.
}
\end{figure}

\begin{table}
\centering
\caption{Comparison of pulsation signals before (Kepler K2) and after (TESS) the MAB, 
grouped by visual inspection (Column 1). 
Columns 2 and 3 list the central peak frequencies in cycles per day and peak amplitudes in parts per thousand (ppt), 
respectively. Columns 4 and 5 provide the signal phase and significance level (Sig)}
 \label{tabMod}
\begin{tabular}{ccccc}
\hline
\multicolumn{5}{c}{\emph{Kepler K2}} \\
\hline
Group & Freq. (d$^{-1}$) & Amp. (ppt) & Phase ($\Theta$) & Sig \\ 
\hline
g0 & 0.4778 & 608 & 2.0452 & 5 \\
g0 & 0.5591 & 678 & 4.4053 & 6 \\
g0 & 0.7854 & 836 & 5.7696 & 8 \\
g0 & 1.3302 & 677 & 3.2917 & 6 \\
g0 & 1.5672 & 633 & 0.1241 & 6 \\
g0 & 1.8628 & 682 & 2.8266 & 6 \\
\noalign{\smallskip}
\textcolor{red}{g1} & \textcolor{red}{2.2732} & \textcolor{red}{1533} & \textcolor{red}{0.7096} & \textcolor{red}{22} \\
\textcolor{red}{g1} & \textcolor{red}{2.3888} & \textcolor{red}{919} &  \textcolor{red}{6.0846} & \textcolor{red}{9} \\
\textcolor{red}{g1} & \textcolor{red}{2.5867} & \textcolor{red}{1142} & \textcolor{red}{1.2655} & \textcolor{red}{14} \\
\textcolor{red}{g1} & \textcolor{red}{2.7130} & \textcolor{red}{841} &  \textcolor{red}{6.1223} & \textcolor{red}{8} \\
g1 & 3.2818 & 690 & 3.5934 & 6 \\
\noalign{\smallskip}
g2 & 4.0769 & 1201 & 2.7526 & 9 \\
\textcolor{red}{g2} & \textcolor{red}{4.3479} & \textcolor{red}{602} & \textcolor{red}{2.7266} & \textcolor{red}{5} \\
g2 & 4.5362 & 939 & 4.0164 & 8 \\
g2 & 4.6092 & 1121 & 0.9218 & 11 \\
\textcolor{red}{g2} & \textcolor{red}{4.8042} & \textcolor{red}{796} & \textcolor{red}{0.2244} & \textcolor{red}{8} \\
g2 & 4.9217 & 1554 & 3.5469 & 23 \\
\hline
\multicolumn{5}{c}{\emph{TESS}} \\
\hline
\textcolor{red}{g1} & \textcolor{red}{2.2743} & \textcolor{red}{646} & \textcolor{red}{5.6860} & \textcolor{red}{266} \\
g1 & 2.3140 & 198 & 4.8089 & 43 \\
\textcolor{red}{g1} & \textcolor{red}{2.3887} & \textcolor{red}{1394} & \textcolor{red}{4.7711} & \textcolor{red}{515} \\
\textcolor{red}{g1} & \textcolor{red}{2.5889} & \textcolor{red}{680} & \textcolor{red}{0.9713} & \textcolor{red}{254} \\
g1 & 2.6535 & 233 & 3.5030 & 53 \\
\textcolor{red}{g1} & \textcolor{red}{2.7107} & \textcolor{red}{202} & \textcolor{red}{1.8395} & \textcolor{red}{39} \\
\noalign{\smallskip}
g2 & 4.2331 & 212 & 1.7646 & 24 \\
\textcolor{red}{g2} & \textcolor{red}{4.3440} & \textcolor{red}{171} & \textcolor{red}{5.6491} & \textcolor{red}{27} \\
g2 & 4.4657 & 368 & 3.3124 & 33 \\
g2 & 4.6346 & 336 & 3.5105 & 36 \\
g2 & 4.7069 & 232 & 0.9903 & 25 \\
g2 & 4.7551 & 321 & 1.9451 & 76 \\
\textcolor{red}{g2} & \textcolor{red}{4.8026} & \textcolor{red}{409} & \textcolor{red}{1.8037} & \textcolor{red}{44} \\
g2 & 4.8741 & 494 & 6.2674 & 161 \\
g2 & 4.9071 & 531 & 2.0142 & 111 \\
g2 & 4.9809 & 1523 & 3.1030 & 444 \\
g2 & 5.0954 & 217 & 4.1239 & 30 \\
\hline
\end{tabular}
\end{table}

The photometric data from ASAS-SN observed the entire of the major outburst (MAB). 
Additionally, the high-precision and high-cadence Kepler K2 data, 
obtained before the MAB, and the TESS light curve corresponding to the new-quiescent (new-QS) state after the MAB, 
provide unique advantages. 
Together, these datasets allow us to investigate the pulsational behavior and its changes before and after the outburst. 

The 36-day baseline Kepler K2 data reveals two minor outbursts on MJD 56775 and 56786, 
with brightness increases of up to 2\% and 5\% respectively, 
separated by 11 days. 
To extract the pure pulsational signal, 
the light curve was detrended by dividing the original data by a 200-point smoothed version. 
Subsequently, the \emph{SigSpec} package was used to analyze the detrended light curve, 
identifying signals with a significance value (Sig) $>$ 5 as real signals, which are listed in Table.~\ref{tabMod}.

On the other hand, the TESS light curve, which corresponds to the new-QS state after the MAB, 
shows 76-day baseline in combination three consecutive Sectors (43, 44, and 45).   
The raw data were visually inspected, and abnormal segments around sector gaps were removed. 
The remaining segments were detrended by dividing the original data by a smoothing spline fit, 
yielding normalized light curves. 
These normalized light curves were then concatenated into a single dataset and analyzed using the \emph{SigSpec} package. 
A total of 66 signals with Sig $>$ 5 were identified, of which 39 were independent frequencies based on the Rayleigh resolution criteria. 
For brevity, Table.~\ref{tabMod} lists the 17 strongest signals with Sig $>$ 20. 
Among these, four high-amplitude signals (highlighted in red) are considered the same pulsation signals as those found in the Kepler K2 data.

Examining outbursts of other Be stars observed by Kepler, such as KIC 9715425, KIC 11953267, and KIC 4939281, 
two common features emerge: mode exhaustion and amplitude variations. 
Mode exhaustion in the target EPIC 202060631 is evidenced by the following: 
As shown in Table.~\ref{tabMod}, the 6 modes in the g0 groups and the peaks at 4.0769 and 4.6092 d$^{-1}$ with amplitudes higher than 1000 ppt observed in the Kepler K2 data significantly decreased in amplitude, 
even falling below the detection limit in the TESS light curve. 
Consequently, the overall power spectrum appears quieter in TESS. 
Although signals at frequencies lower than 1 d$^{-1}$ may be affected by the outburst and detrending issues, 
the amplitude decrease for peaks around 4 d$^{-1}$ is evident.

On the other hand, amplitude variations are revealed in the ``common modes'' highlighted in red between Kepler K2 and TESS data. 
Both the absolute amplitudes and relative intensities among these modes changed before and after the major outburst.

Furthermore, amplitude variations may also be present in the ASAS-SN photometric data. 
To study short-timescale variations, the original ASAS-SN light curve was divided by a 15-point smoothed version to obtain a residual light curve. 
The results show a standard deviation of about 0.7 percent in the residuals, 
which tripled within 150 days after the outburst's onset. 
Although no periodic signals were detected in the ASAS-SN data, 
the increased magnitudes of the residuals likely point to a real physical amplification of the stellar pulsations, rather than merely a visual effect.

Considering that pulsation amplitudes were enhanced in both pole-on (KIC 9715425) and edge-on (KIC 11971405) stars shown in Kepler data, the enhancement may be isotropic rather than anisotropic and global rather than localized. 
This serendipitously agrees with the radial velocity variations observed in the helium lines studied in the next section.

\section{The movement of stellar photosphere indicated by Helium lines}

\begin{figure*}
\includegraphics[width=0.7\linewidth]{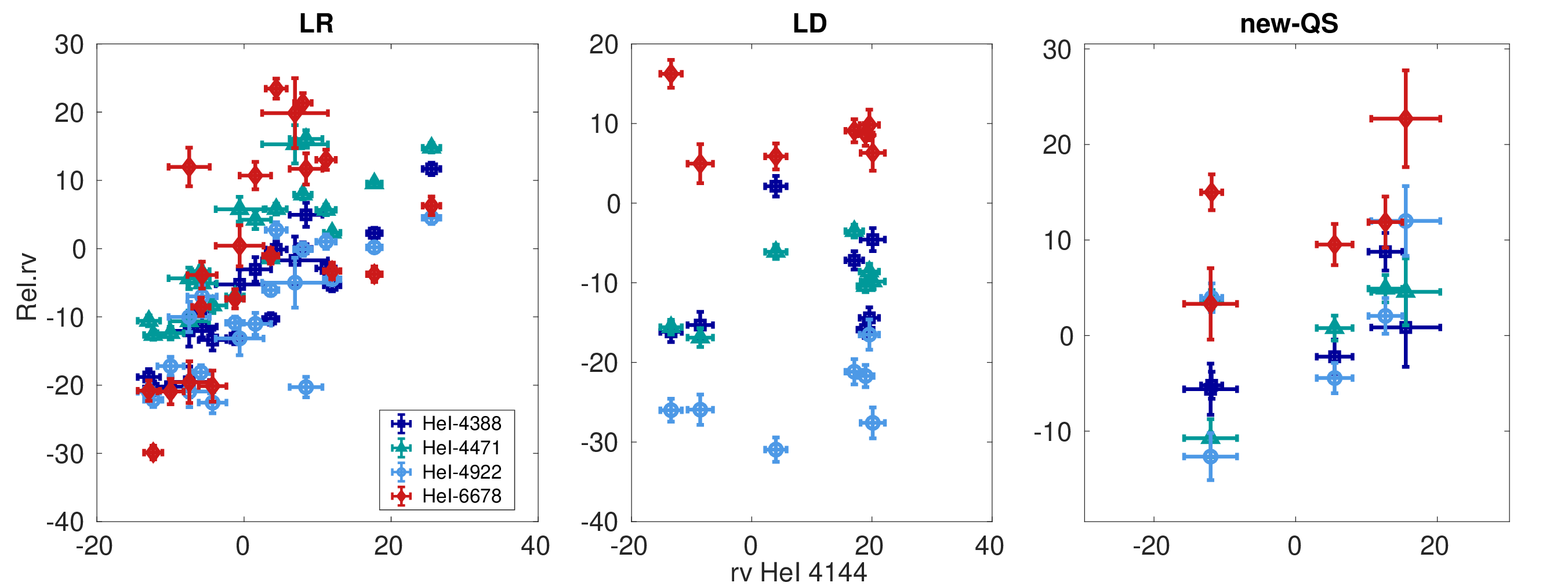}
\caption{\label{HervsCor} The relation shape on Radial Velocities (RVs) between He\,{\sc i} 4144 and the others 4 Helium absorption lines. 
The relation with Helium at 4388, 4471, 4922 and 6678 were shown in dark blue squre, 
cyan triangle, light blue open circle, red diamond respectively. 
Pronounced positive both shown in LR and new-QS phase, while in LD phase the correlation was missing.  
}
\end{figure*}

\begin{figure}
\includegraphics[width=1.0\linewidth]{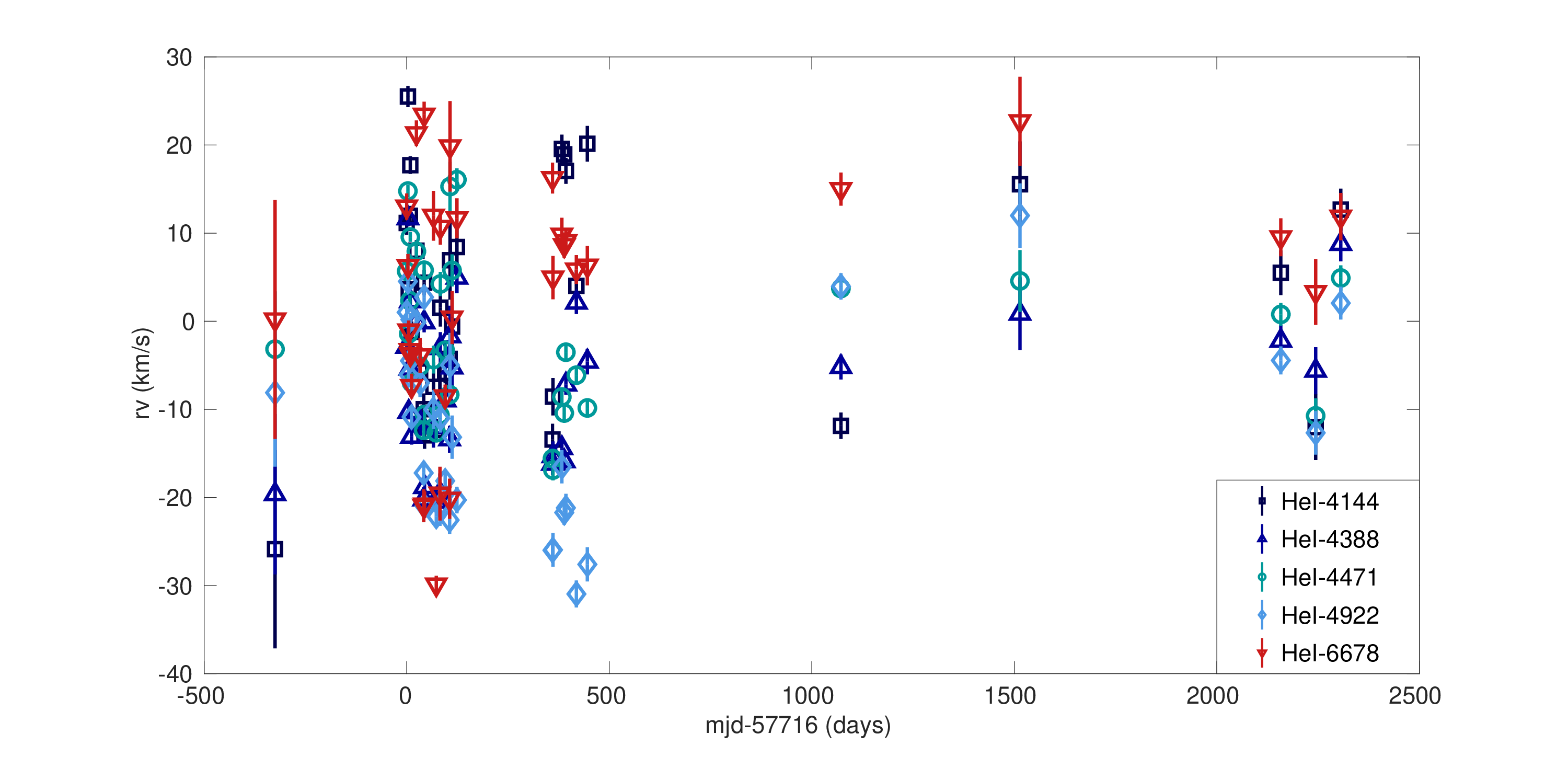}
\caption{\label{Hervs} The relative Radial Velocities (RVs) variation with time for He\,{\sc i} center at 4144, 4388, 4471, 4922, and 6678. 
The RVs in LR general have blue shift trend. 
The difference of RV approached at maximum in LD phase, e.g. when He\,{\sc i} 6678 (red reverse triangle) in red shift but He\,{\sc i} 4922 (light blue diamond) 
in the blue shift region. 
While the target brightness fallen back to quiescent, especially the last 3 observations, the RVs closed to each other, even agreement in error. 
}
\end{figure}

\begin{figure*}
\includegraphics[width=1.0\linewidth]{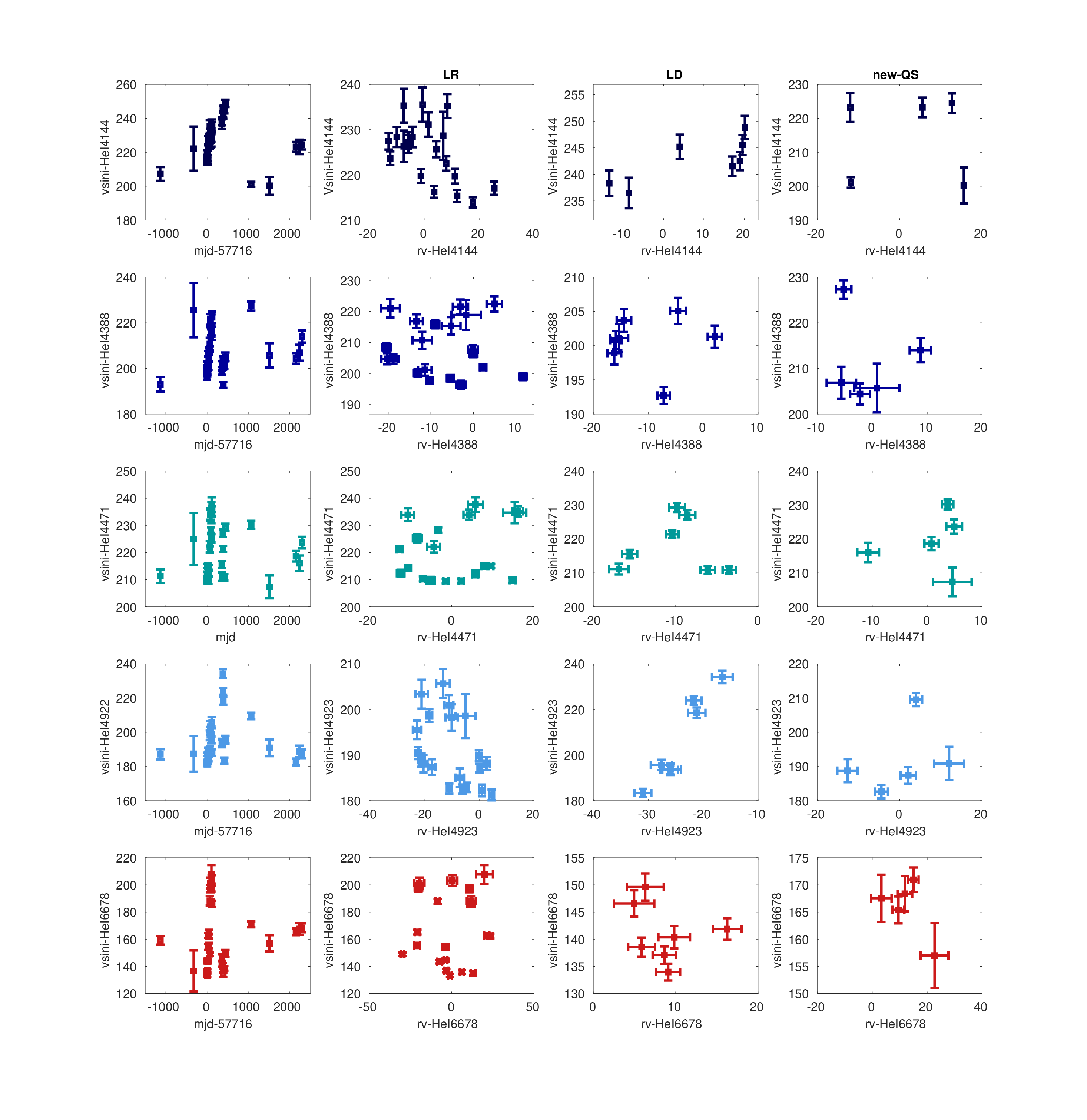}
\caption{\label{Hervvsini} Relationship between Radial Velocities (RVs) and 
projected rotational velocity ($v$sin\,$i$) for the He\,{\sc i} 4144, 4388, 4471, 4922, and 6678 lines, 
shown in black, dark blue, cyan, light blue, and red error bars, respectively, from top to bottom. 
The first column displays the $v$sin\,$i$ variation with time for each line, 
exhibiting higher values during the outburst phase compared to the quiescent (QS) phase. 
The second, third, and fourth columns illustrate the RV-$v$sin\,$i$ relationship for each line during different outburst stages.
}
\end{figure*}

\begin{figure}
\includegraphics[width=0.85\linewidth]{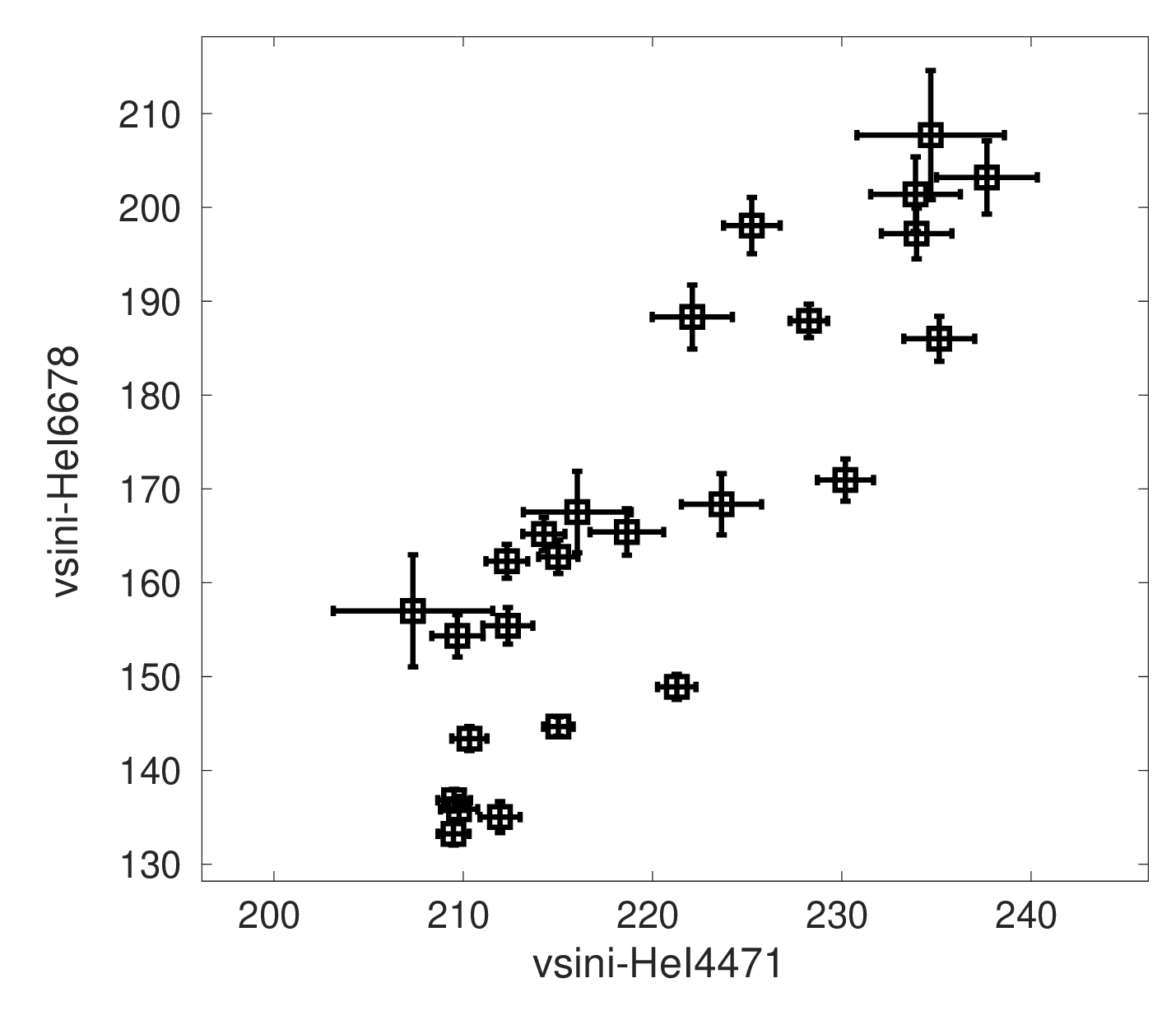}
\caption{\label{He7178} Relationship between vsini for He\,{\sc i} 4471 and He\,{\sc i} 6678 during the LR and new-QS phases. 
The $v$sin\,$i$ values were calculated from FWHM divided by the central wavelength and multiplied by the speed of light.
}
\end{figure}

\begin{table}
\centering
\caption{\label{tabRvs} Radial velocity variations in km\,s$^{-1}$ derived from template fitting of photospheric absorption lines (absRV), 
and centroid velocities of H$\alpha$ (H$\alpha$\,RV) and H$\beta$ (H$\beta$\,RV) emission lines. 
All velocities are normalized to the first spectrum (ID=1). 
Note that the H$\beta$ line is absent in the second spectrum (ID=2), 
and the H$\alpha$ region is missing for the 30th spectrum (ID=30) for unknown reasons.} 
\begin{tabular}{ccccc}
\hline
ID & MJD & abs\,RV & H$\alpha$\,RV & H$\beta$\,RV \\
\hline
2 & 57391 & $-10.9\pm1.1$ & $-27.9\pm0.5$ & - \\
3 & 57716 & $3.2\pm0.9$ & $-18.1\pm0.3$ & $-10.0\pm0.7$ \\
4 & 57719 & $13.1\pm0.8$ & $-14.2\pm0.2$ & $2.3\pm0.6$ \\
5 & 57721 & $-3.6\pm0.9$ & $-20.7\pm0.2$ & $-18.3\pm0.6$ \\
6 & 57724 & $0.2\pm0.9$ & $-21.2\pm0.2$ & $-11.6\pm0.6$ \\
7 & 57725 & $8.2\pm0.8$ & $-16.4\pm0.2$ & $-1.4\pm0.5$ \\
8 & 57728 & $-9.1\pm0.8$ & $-21.8\pm0.3$ & $-13.6\pm0.7$ \\
9 & 57740 & $3.2\pm0.9$ & $-26.0\pm0.2$ & $-30.3\pm0.7$ \\
10 & 57749 & $-8.5\pm1.2$ & $-22.6\pm0.4$ & $-15.8\pm1.2$ \\
11 & 57758 & $-16.2\pm0.9$ & $-20.2\pm0.3$ & $-17.0\pm0.9$ \\
12 & 57759 & $2.3\pm0.9$ & $-30.8\pm0.3$ & $-32.4\pm0.9$ \\
13 & 57760 & $-16.4\pm0.9$ & $-19.3\pm0.3$ & $-5.6\pm0.8$ \\
14 & 57782 & $-11.1\pm1.1$ & $-32.2\pm0.4$ & $-52.5\pm2.0$ \\
15 & 57789 & $-20.3\pm0.9$ & $-25.2\pm0.2$ & $-16.9\pm0.7$ \\
16 & 57798 & $-18.0\pm1.0$ & $-29.7\pm1.4$ & $-29.4\pm1.7$ \\
17 & 57799 & $-5.5\pm0.9$ & $-11.6\pm0.3$ & $8.0\pm1.3$ \\
18 & 57811 & $-13.8\pm2.1$ & $-25.2\pm0.1$ & $-6.3\pm0.6$ \\
19 & 57822 & $-15.9\pm0.9$ & $-15.0\pm0.1$ & $-8.0\pm0.9$ \\
20 & 57823 & $1.5\pm1.6$ & $-30.2\pm0.2$ & $-29.3\pm2.6$ \\
21 & 57828 & $-7.8\pm0.9$ & $-36.9\pm0.0$ & $-31.2\pm1.5$ \\
22 & 57840 & $-3.4\pm1.1$ & $-19.2\pm0.4$ & $-4.6\pm0.8$ \\
23 & 58076 & $-23.1\pm0.8$ & $-7.1\pm0.1$ & $-21.7\pm0.4$ \\
24 & 58077 & $-21.6\pm0.9$ & $-6.2\pm0.2$ & $-17.9\pm0.6$ \\
25 & 58099 & $-10.1\pm0.8$ & $-10.9\pm0.1$ & $-20.3\pm0.5$ \\
26 & 58105 & $-13.8\pm0.8$ & $-12.0\pm0.2$ & $-21.2\pm0.4$ \\
27 & 58109 & $-6.4\pm0.8$ & $-10.7\pm0.2$ & $-12.1\pm0.4$ \\
28 & 58135 & $-11.8\pm1.0$ & $-9.6\pm0.1$ & $-7.1\pm0.4$ \\
29 & 58162 & $0.1\pm1.5$ & $-19.1\pm0.1$ & $-11.4\pm0.5$ \\
30 & 58788 & $-4.9\pm0.9$ & - & $9.1\pm0.4$ \\
31 & 59230 & $3.8\pm1.4$ & $8.2\pm0.3$ & $15.3\pm1.5$ \\
32 & 59874 & $0.9\pm1.4$ & $-11.2\pm0.2$ & $-8.3\pm1.0$ \\
33 & 59960 & $-11.0\pm1.1$ & $-12.9\pm0.4$ & $-11.8\pm2.1$ \\
34 & 60022 & $5.9\pm1.0$ & $-4.5\pm0.3$ & $11.9\pm1.8$ \\
\hline
\end{tabular}
\end{table}

For a B-type star with an effective temperature of around 18000 K, multiple helium absorption lines are expected to be present in the observed spectrum, originating primarily from the photospheric layers. Visual inspection revealed prominent He\,{\sc i} absorptiong lines at 4009, 4026, 4144, 4388, 4471, 4922, 5018, 5875 , 6678 and 7065 \AA. 
To balance complexity we select five lines: 4144, 4388, 4471, 4922 and 6678 for further analysis. 
Traditionally, the circumstellar continuum in the optical wavelength range (4000-8000 \AA) is expected to exhibit a power-law like shape. 
Therefore, on the scale of individual spectra lines, the influence of circumstellar continuum can be approximate as a constant value.

Consequently, the central wavelengths and full widths at half maximum (FWHM) of these lines should not be affected by the circumstellar continuum.
The central wavelengths are related to the radial velocity and radial motion of the star photosphere, 
while the FWHM values are indicators of the projected rotational velocity and tangential motion. 
We employed Gaussian fitting to derive the central wavelengths and FWHM for each line. 
Based on the ASAS-SN photometric observations of the outburst, we divided the LAMOST spectra into three groups: the light rising (LR) phase for number 3-22 , the light decay (LD) phase from 23-29, and the new-quiescent (new-QS) phase after the outburst.  
The central wavelengths of these helium lines in the first LAMOST spectrum, obtained during the quiescent period before the outburst  (MJD at 56583), were set as the zero-point for radial velocities (RVs) calculations. 

In analyzing the spectral data of this Be star during different outburst stages, we have encountered some thought-provoking phenomena. 
Firstly, the radial velocity variations of different helium absorption lines showed good correlation during the LRand the new-QS stages, but this correlation disappeared during the LD stage (see Fig.~\ref{HervsCor}). 
Secondly, almost all lines exhibited general a blueshift trend with significant dispersion in the LR stage, 
their radial velocities scattered to different positions during the LD stage, with some showing redshifts and others showing blueshifts. 
Finally, when the star's brightness returned to the quiescent state, the radial velocities of these lines converged again (see Fig.~\ref{Hervs}).

This phenomenon suggests that the motion state of the photosphere underwent dramatic changes during the outburst process, which might be closely related to the internal physical processes of the star. We have ruled out the possibility of a simple binary system and the influence of the continuous spectrum.
However, we cannot discount the possibility that the optically thick circumstellar disk obscures the shape of the absorption line or contaminates it with anonymous emissions within the absorption trough.

During the outburst activity, the violent energy transportation and material motions inside Be stars may excite some peculiar non-radial pulsation modes that are less detectable during quiescent periods. These abnormal pulsation modes could cause dramatic disturbances in the motion state of the photosphere, leading to prominent differences in the kinematics of different atmospheric layers and ionization regions where the spectral lines originate. Consequently, this may result in the observed dispersion and lack of coordination in the radial velocity variations of different spectral lines.  This interpretation is consistent with the Kepler observations of increased pulsation amplitudes in Be stars during outbursts e.g. KIC9715425, KIC6954726, KIC11953267, KIC4939281. Considering the standard deviation of those RVs about 12 km\,s$^{-1}$ just only a little higher than standard pulsation velocities.

Analysis of the spectral line profiles revealed an intriguing relationship between the 
radial velocity (RV) and projected rotational velocity ($v$sin\,$i$) 
variations during different outburst phases. 
First, the $v$sin\,$i$ values were calculated from FWHM divided by the central wavelength and multiplied by the speed of light, 
which yielded values consistent with those obtained from the UlySS fitting. 
Interestingly, for some lines like He\,{\sc i} 4144, 4388, 4923, there was a clear positive correlation between RV and $v$sin\,$i$ 
during the LR phase, while this correlation became negative during the LD phase (see Fig.~\ref{Hervvsini}). 
Although the RV-$v$sin\,$i$ relationship for He\,{\sc i} 4471 and 6678 was poor in all phases, 
Fig.~\ref{He7178} reveals a striking positive correlation between their $v$sin\,$i$ values. 
Considering that these two lines are located at the red (He\,{\sc i} 6678) and blue (He\,{\sc i} 4471) 
extremes of the LAMOST spectra range, artifacts due to wavelength calibration can be ruled out. 
This suggests that the photosphere underwent dramatic changes during the outburst, 
such as rotational acceleration or expansion, which are both plausible scenarios.

\section{Disk kinematics variation with proxy of Hydrogen and Metallic emission lines}

\subsection{Disk density profile variation studied by HDUST}

\begin{figure*}
\includegraphics[width=1.0\linewidth]{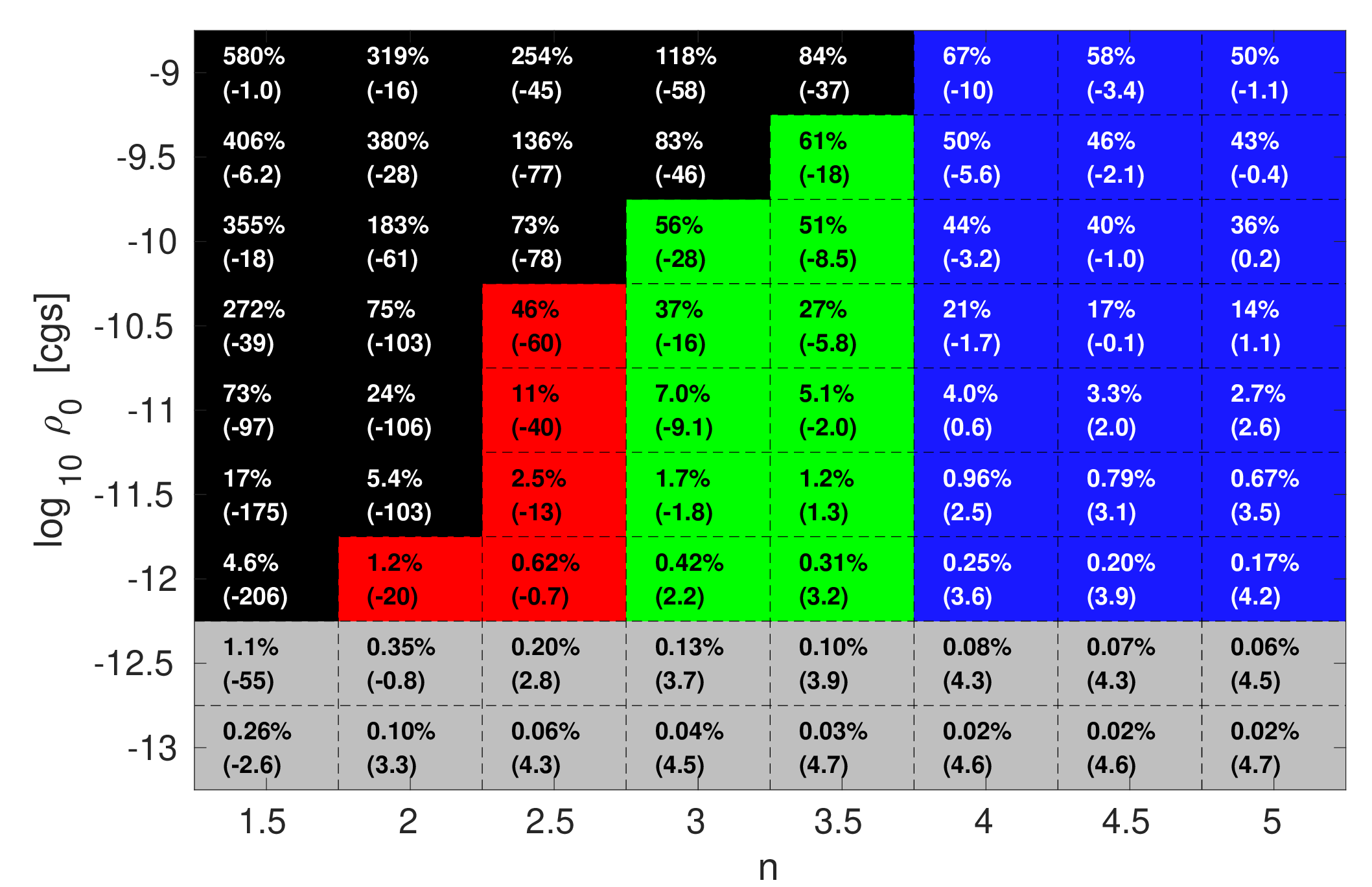}
\caption{\label{dVEW} The V-band brightness brightening percentage predicted in each model and equivalenth width (EW)
of H$\alpha$ emission predicted in each model are labeled into 1st and 2nd line respectively. 
The five different regions which includes disk formating, steady-state disk, disk-dissipating, detection limit and forbidden region were colored in background with blue, green ,red, grey and black color.
}
\end{figure*}

\begin{figure}
\includegraphics[width=1.0\linewidth]{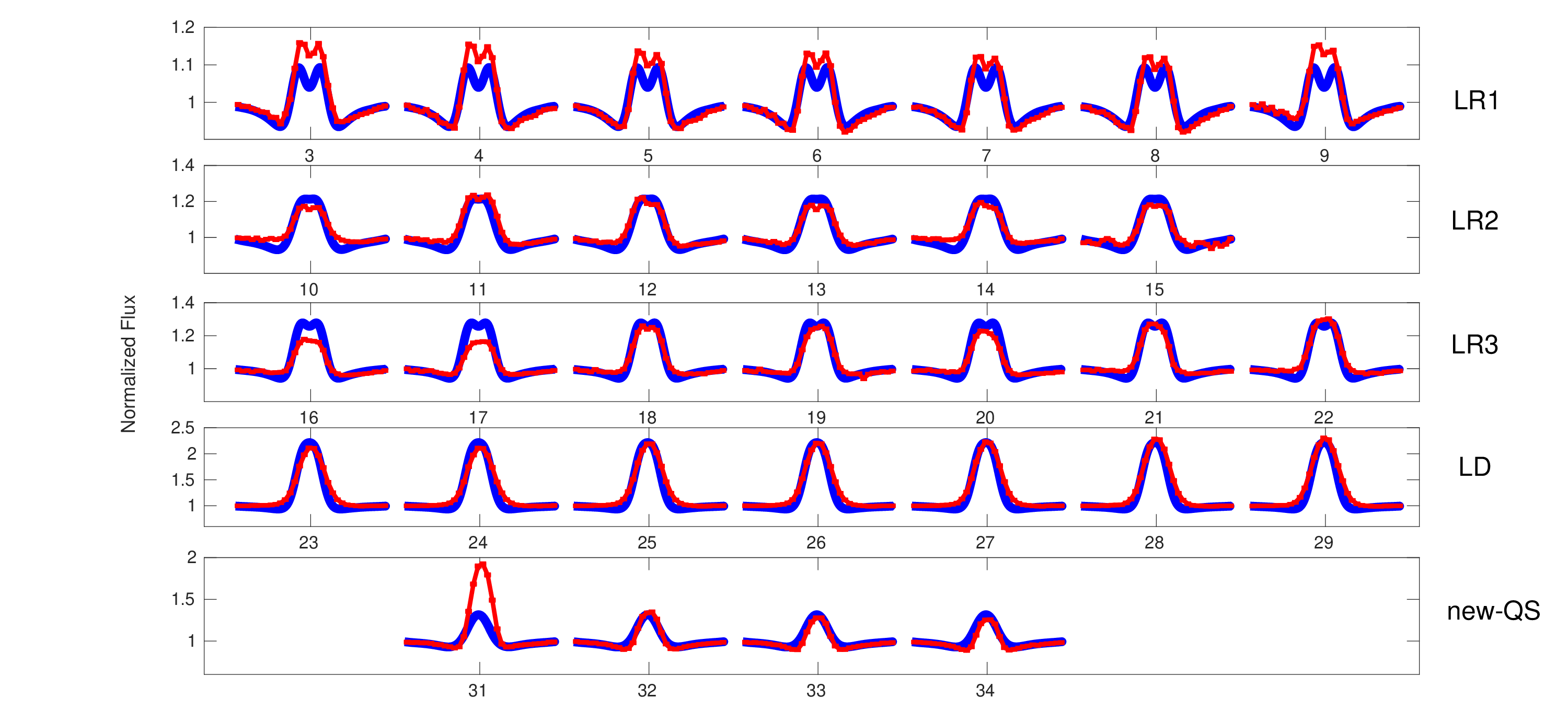}
\caption{\label{HaHDUST} Comparison between LAMOST low-resolution H$\alpha$ emission and the best-matched HDUST models for the 5 different outburst periods. 
The thick blue lines represent the best-matched models with specific parameters listed in Table 6.1.1. 
The red point-lines are observation spectra, with the observation numbers listed at the bottom. 
Due to the diversity of real emission variations, the rough classification may not provide the best explanation, e.g., ID=30 (without published observation for unknown reason) and number 31. The spectra are normalized to the continuum flux, and the state names (LR1, LR2, LR3, LD, new-QS) are labeled on the right. 
}
\end{figure}

\begin{figure}
\includegraphics[width=1.0\linewidth]{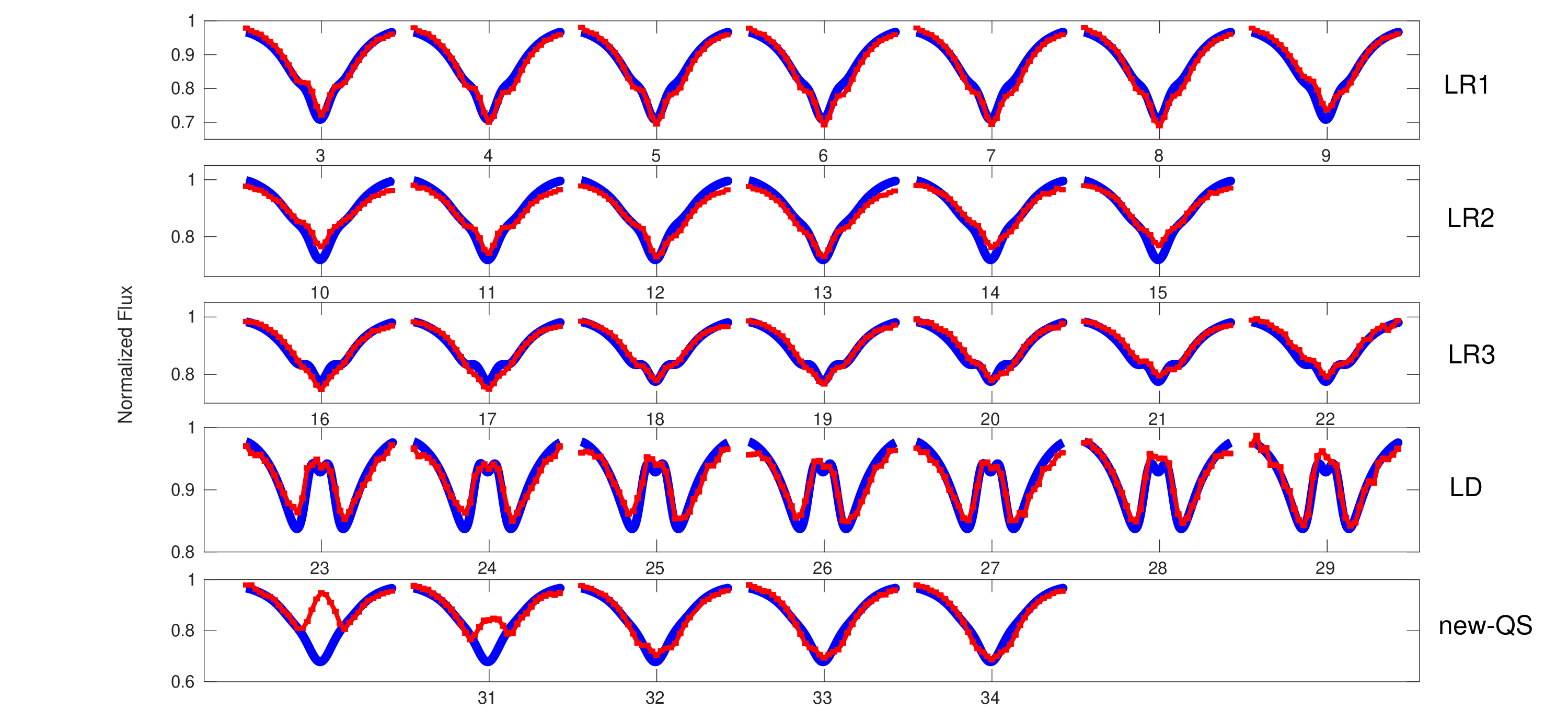}
\caption{\label{HbHDUST} The same as Fig.~\ref{HaHDUST} but for H$\beta$.
}
\end{figure}

\begin{figure}
\includegraphics[width=1.0\linewidth]{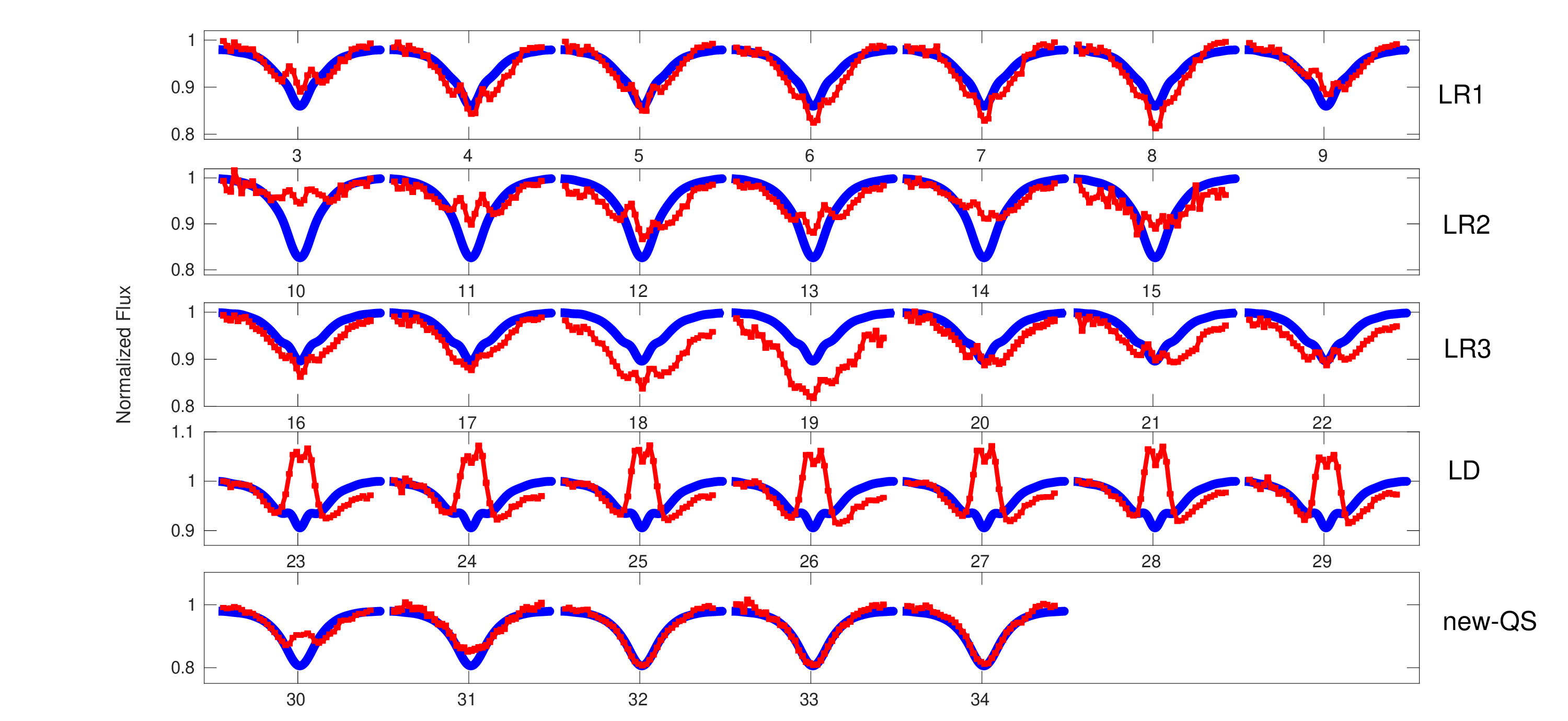}
\caption{\label{Pa14HDUST} The same as Fig.~\ref{HaHDUST} but for Paschen 14 center at 8600 \AA.
}
\end{figure}

\begin{table}
\centering
\caption{\label{tabHDUST} Best-matched HDUST disk model parameters (log$_{10}$ $\rho$0 and $n$) for different outburst periods, 
along with the corresponding V-band brightness enhancement due to the disk. 
The LR period is visually classified into three sub-stages (LR1, LR2, LR3), 
while the LD and new-QS periods are also listed in the first column.}
\begin{tabular}{cccc}
\hline
Period & $n$ & $\log_{10}\rho_0$ & $\Delta V$ \\
\hline
LR1 & 4.5 & -10.5 & 17\% \\
LR2 & 2.5 & -12.0 & 0.62\% \\
LR3 & 4.0 & -10.5 & 21\% \\
LD & 3.0 & -11.0 & 7\% \\
new-QS & 2.0 & -12.5 & 0.2\% \\
\hline
\end{tabular}
\end{table}

According to the Viscous Decretion disk (VDD) models 
(Carciofi \& Bjorkman 2006, 2008), 
the circumstellar disk density profile can be simplified as a power-law function controlled by 2 parameters: 
the disk base density $\rho$0 (in cgs units) and the power-law index $n$ for the volume density variation (see 
Vieira et al. (2017)
Section 2). 
Base on the volume density index -- $n$  value, the disk state can be categorized into 3 regions: 
forming disks, steady-state disks, and dissipating disks. When log$_{10}$$\rho$0 $<$ -12.5, 
the disk brightening is barely detectable in both optical and infrared wavelengths, defining the detection limit region.  
The complete classification is illustrated in Figure 13 of the work of Vieira et al. (2017)

For the EPIC202060631 system, I fixed the central star parameters provided by BRUCE04. 
The stellar parameters were chosen with $n$ ranging from 1.5 to 5 in steps of 0.5, and log$_{10}$$\rho$0 [cgs] ranging from -13 to -9 in steps of 0.5, resulting in a total of 72 models. 
The results include the disk Spectral Energy Distribution (SED) and emission line profiles for each model run. Consequently, the corresponding percentage increase in V-band brightness and the predicted equivalent width of the
H$\alpha$ line for each disk model are plotted in Fig.~\ref{dVEW}. 
Generally, a disk with lower $n$ and lower base density trends to exhibit stronger emission, 
while a disk with higher base density is more likely to have a larger V-band brightening.

To study the disk characteristics for each period of the EPIC202060631 outburst, 
I compared the H$\alpha$, H$\beta$, and Pa14 (center at 8600 \AA) line profiles from the above models with the LAMOST low-resolution spectra obtained during different outburst periods. 
In addition to the spectra classification based on different periods discussed in Section 5, 
where number 23-29 wer in the LD phase and 30-34 were in the new-QS phase, 
I visually divided the LR period into 3 sub-groups: number 3-9 as LR1, 10-15 as LR2, and 16-22 as LR3. 
The $n$ and log$_{10}$$\rho$0 parameters of the best-matching model for each state, 
as well as the corresponding V-band light increase percentage, are listed in Table.~\ref{tabHDUST}.

For these 5 difference stages, I compared the observed H$\alpha$, H$\beta$, and Pa14 lines with the models in 
Figs.~\ref{HaHDUST}, \ref{HbHDUST}, and \ref{Pa14HDUST}. 
It can be seen that the Pa14 model deviates more from the observations compare to H$\alpha$ and H$\beta$. 
This suggests that the disk structure may be more complex than that simplified model assumption. 
The model predictions for the V-band brightening are quite close to the observed brightening. 
However, during the LR2 stage, the emission line state is closer to a dissipating disk, 
but a dissipating disk typically corresponds to a smaller V-band brightening, less than 1\%, 
which is lower than the observed brightening. 
This further indicates that the cause of the V-band brightness variation is complex, 
and changes in the stellar photosphere may also have occurred. 
Moreover, the transition form a steeper (higher $n$ index) to a flatter 
(lower $n$ index) disk also suggests that multiple mass injection events 
from the stellar photosphere into the circumstellar disk may have occurred during the LR period of the entire outburst.

\subsection{The complex RV movement by H$\alpha$ and H$\beta$ emission profile}

\begin{figure}
\includegraphics[width=1.0\linewidth]{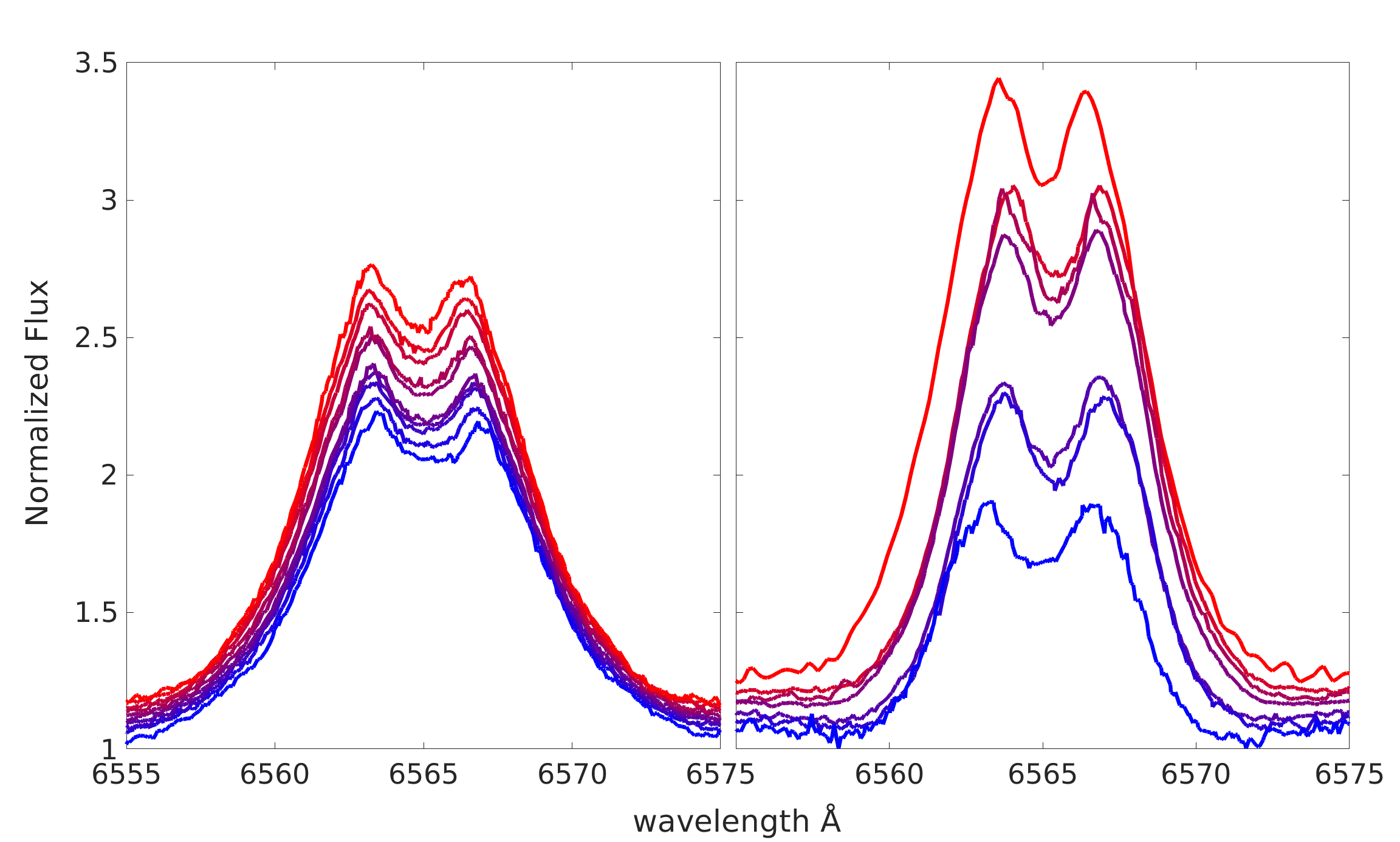}
\caption{\label{MedHa} Evolution of the H$\alpha$ line centroid velocities derived from medium-resolution (R $\sim$ 7200) spectra. 
The emission intensity peaked at MJD 58559, shown as the highest red spectrum in the right panel. 
Spectra before this maximum are plotted in the left panel with a positive vertical offset of 0.015 for clarity. 
As the emission intensity increases with time, the line profiles transition from blue to red. 
Conversely, spectra after the maximum (right panel) are offset negatively and transition from red to blue as intensity decreases. 
The y-axis scale is kept consistent across panels for easy comparison.
}
\end{figure}

\begin{figure*}
\includegraphics[width=1.0\linewidth]{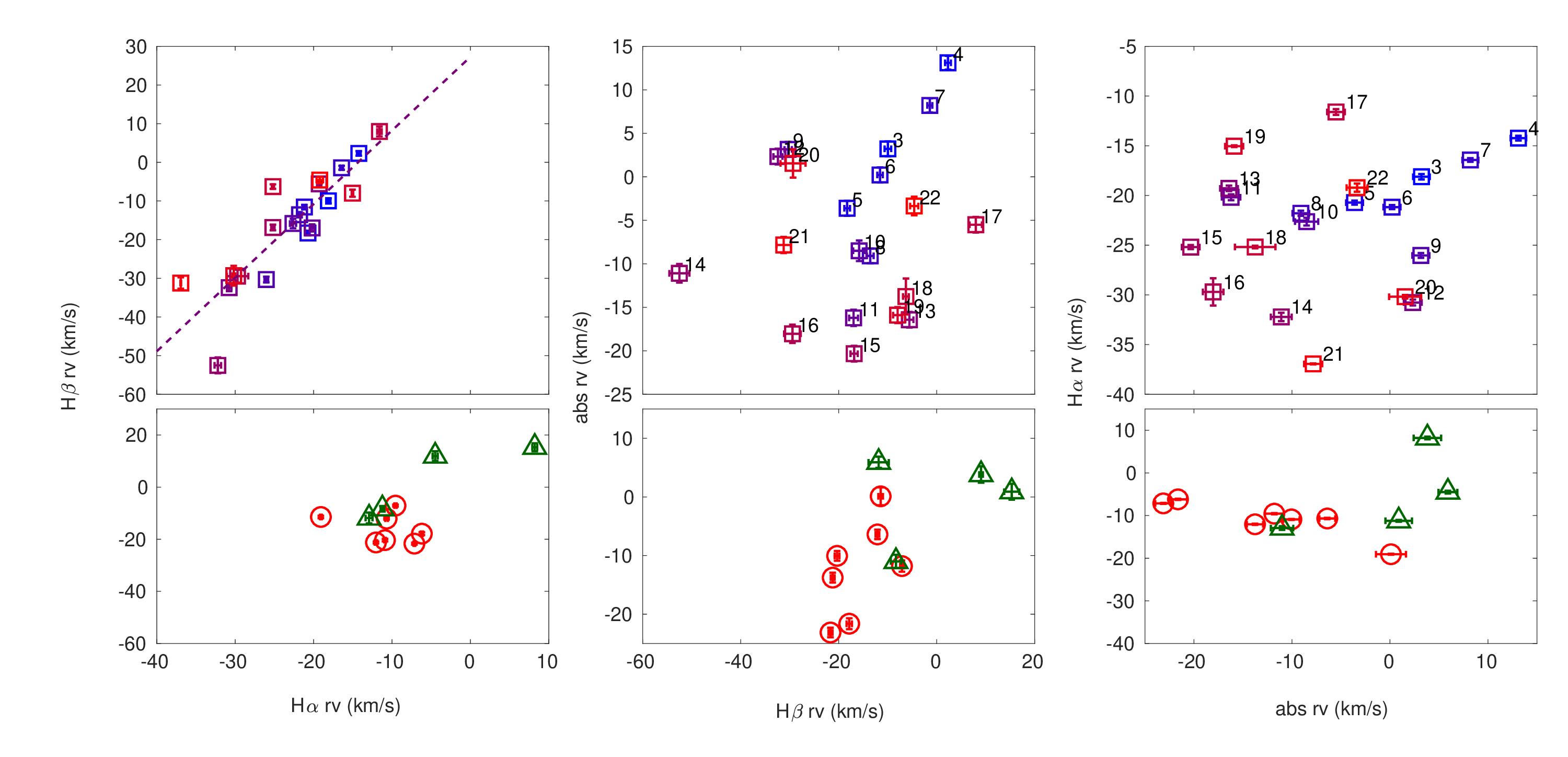}
\caption{\label{EmiRvs} Correlations between the centroid radial velocities (RVs) of H$\alpha$, H$\beta$, and photospheric absorption lines (abs RV) 
during different outburst periods. 
Top three panels: LR period, with data points color-coded from blue to red with time progression. 
A tight linear correlation exists between H$\alpha$ and H$\beta$ RVs (purple dashed line). 
Observation IDs are labeled in the top-right panels. 
Bottom three panels: LD period (red circles) and new-QS period (green triangles). 
During LD, H$\alpha$ RVs exhibit a general blueshift trend consistent with medium-resolution spectra, 
while H$\beta$ RVs show a possible redshift, indicating a negative correlation. 
In the new-QS period, RVs for both lines appear chaotic, also in agreement with medium-resolution spectra.
}
\end{figure*}

In addition to the changes in emission profiles, the centroid radial velocities (RV) also exhibited significant variations throughout the outburst periods. 
Notably, the H$\alpha$ and H$\beta$ emission profiles appeared to move in unison. 
To illustrate these centroid movements intuitively, Fig.~\ref{MedHa} shows medium-resolution spectra. 
During the Light Decay (LD) period, shown in the left panel, the H$\alpha$ emission intensities increased monotonically with an evident overall blueshift trend. In contrast, during the new-quiescent (new-QS) period, the centroids fluctuated chaotically.

Given the broader time coverage of the low-resolution spectra, 
I calculated the centriods of the H$\alpha$ and H$\beta$ emission profiles for each observation. 
There were a total of 20 observations during the LR, 7 during the LD, and 5 during the new-QS period. 
First, the absorption valley from the stellar photosphere was subtracted from the emission profiles. 
However, double-peaked structures persisted at the center. 
Therefore, Gaussian fitting was performed only on the outer contours to determine the central wavelength positions of the emission lines. 
The radial velocity (RV) values were then derived by setting the zero-point at the absorption line center of the first spectrum obtained furing the complete quiescent 
state (QS) period. For comparison, Fig.~\ref{EmiRvs} also shows the relative RV variations of the absorption feature, which serves as a proxy for photospheric variations.

H$\alpha$ and H$\beta$ emission line radial velocities exhibit a remarkably tight linear correlation although the dispersion in 
H$\beta$ is larger than H$\alpha$. 
Physically, during the LR, the circumstellar disk regions probed by H$\alpha$ and H$\beta$ emissions are in close proximity, 
displaying velocity coherence that may indicate an overall expansion of the disk. 
In LD, H$\beta$ traces the inner parts relative to H$\alpha$. 
Thus, the divergence between H$\alpha$ and H$\beta$ could signify ongoing expansion in the outer disk but infalling materical in interior regions.

In contrast, there is no apparent correlation between the radial velocities of the emissions and the absorptions (abs RV). 
The chaotic behavior implies a complex relation between the stellar photosphere and the disk. 

The poor RVs correlations between absorption and emission lines, 
which may trace the photospheric and immediate circumstellar enviroment of the Be star, 
suggest that the disk evolution is decoupled from photospheric changes. 
This disfavors disk obscuration as the cause for the helium line RVs variations, 
since obscuration should produce anti-correlated velocity shifts between disk emission and photospheric absorption. 
A binary origin is also disfavored, as emission from an accretion disk around a companion would be anti-correlated with the helium line RVs. 
However, we cannot rule out that the observed helium line radial velocity and project rotation variations could stem from changing stellar pulsation modes 
or additional unknown absorption or emission components during the outburst, 
since we lack detailed spectroscopic profiling from the low-resolution data.
None periodic signal is detected for either of above RVs, 
not only for any single period but also throughout the entire observation. 
The general redshift trend may indicate that large part of the disk has fallen back onto the stellar surface by the end.

\subsection{H$\alpha$ emission flux intensity vs. EWs}

\begin{figure}
\includegraphics[width=1.0\linewidth]{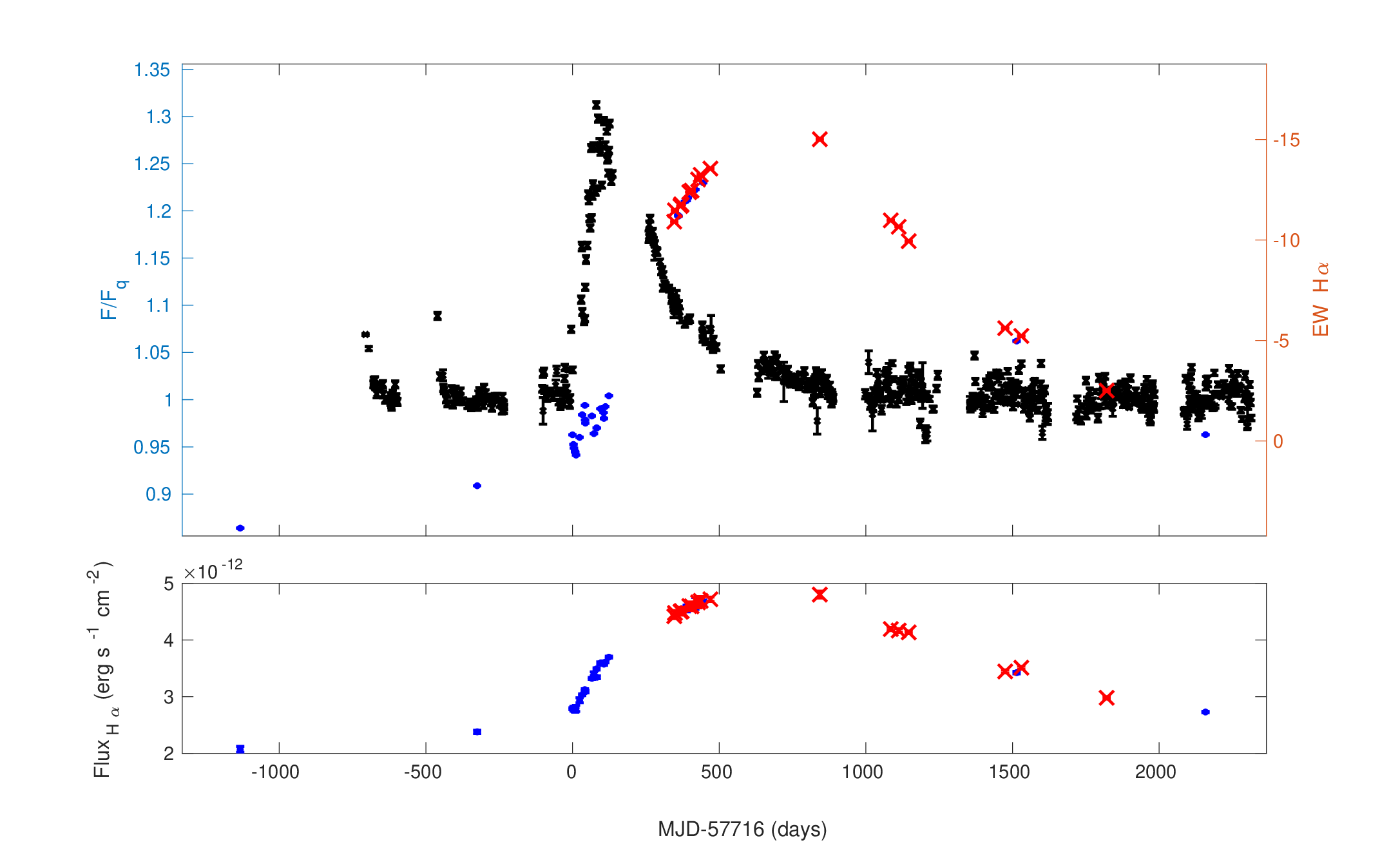}
\caption{\label{EmiHa} Top panel: Comparison of the ASAS-SN V-band relative flux with the variations in the H$\alpha$ equivalent widths (EWs) over time. 
EWs measured from low-resolution spectra are plotted as blue error bars, 
while those from medium-resolution spectra are shown in red. 
Bottom panel: The same data, but displaying the absolute H$\alpha$ flux intensity estimated from the V-band flux and the spectra.
}
\end{figure}

As illustrated in the top panel of Fig.~\ref{EmiHa}, 
the H$\alpha$ EWs exhibit noticeable oscillations during the LR state, 
despite the overall increasing trend. 
However, the EW represents the relative strength of the emission line against the continuum. 
An EW does not necessarily imply a weaker emission but could result from an overall stronger continuum. 
Therefore, the absolute fluxes in the H$\alpha$ region should be calculated based on the H$\alpha$ EWs, 
V-band fluxes, and model predictions.

The LAMOST observations encompass both low and medium-resolution spectra. 
Since there were no contemporaneous observations between the photometric and LAMOST observations, 
the V-band fluxes at the LAMOST observation epochs were calculated through linear interpolation of the ASAS-SN data points. 
The wavelength range from 6556 to 6574 \AA was chosen as the H$\alpha$ flux intensity integration region, 
fully covering the emission profile. 
The continuum flux level in the H$\alpha$ region during the QS was predicted using the BRUCE04 model (see Section 4). 
Subsequently, the continuum flux in the H$\alpha$ region for subsequent observations was assumed to be proportional to the predicted V-band flux variations. 
It should be noted that the integration still includes the absorption part from the central star, 
which is challenging to separate, especially for medium-resolution spectra. 

The result is evident, as shown in the bottom panel of Fig.~\ref{EmiHa}. 
The absolute H$\alpha$ flux intensity increased monotonically during the LR phase. 
However, the intensity reached its maximum at Day 840 relative to MJD at 57716, lagging behind the V-band peak brightness by 760 days. 
That are consistent between the EW measurements and the H$\alpha$ flux intensities.

\subsection{Multiple Mass Ejections Suggested by Balmer Decrements -- D34 and D54}

\begin{figure}
\includegraphics[width=1.0\linewidth]{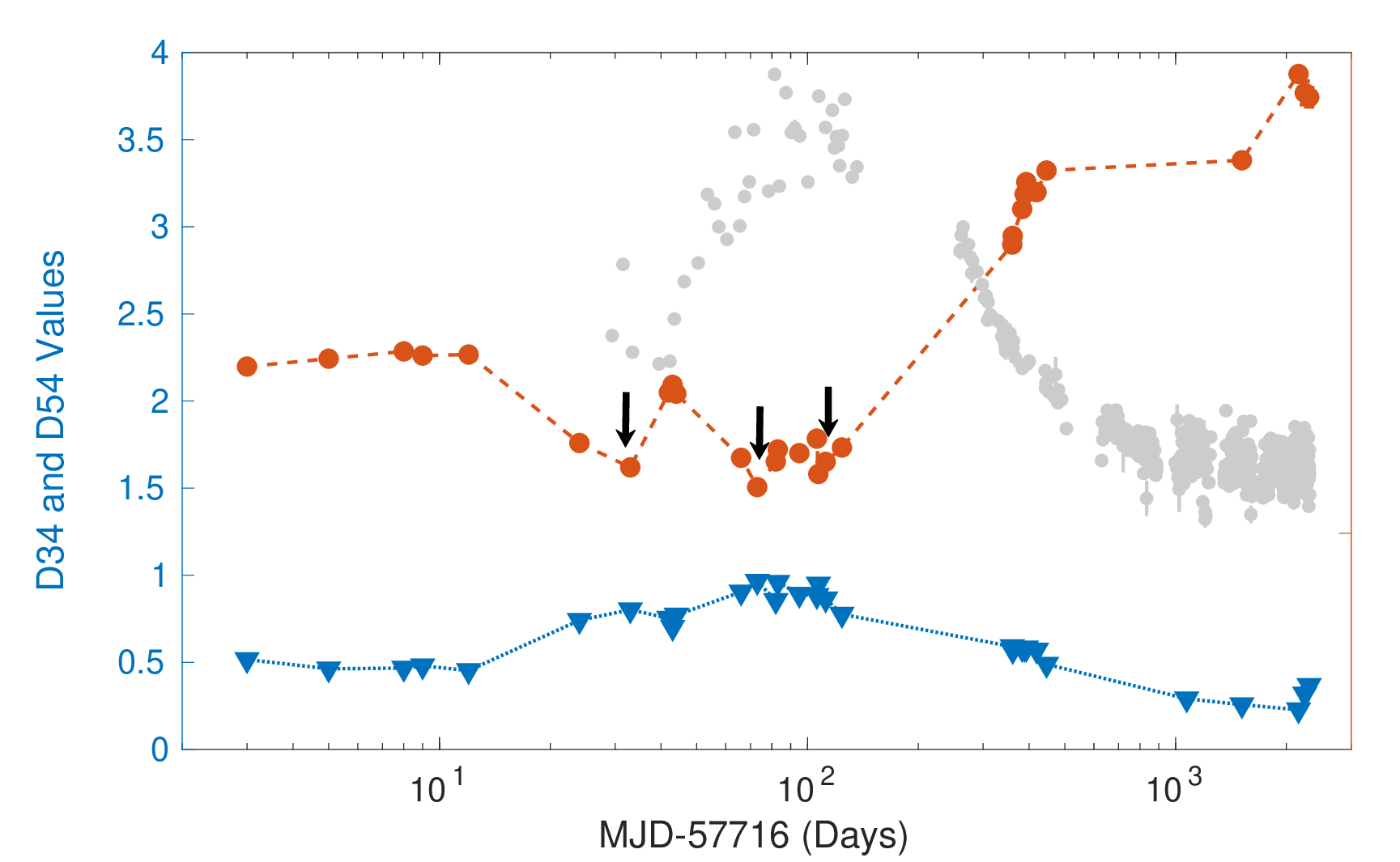}
\caption{\label{D34D54} Variations of diffuse interstellar band ratios D34 and D54 compared to ASAS-SN visual light curve. 
Red open circles connected with dotted lines represent time-resolved D34 ratio. 
Blue open triangles connected with dotted lines denote time-resolved D54 ratio. 
Gray data points with error bars in background display the ASAS-SN light curve over the same period for comparison. 
To accentuate the variations of D34 and D54 during brightening, a logarithmic scale is adopted for the x-axis. 
Black arrows indicate the local minimums of D34.
}
\end{figure}

The Balmer decrement, defined as the ratios of the H$\alpha$ and H$\gamma$ line fluxes to H$\beta$ ($\rm D34=F(H\alpha)/F(H\beta)$ 
and $\rm D54=F(H\gamma)/F(H\beta))$, serves as a useful diagnostic for the electron density and optical depth of 
the circumstellr disk around Be stars. 
Theoretically, for optically thin, low-density Case B recombination conditions, D34 $>=$ 2.7 and D54 $<=$ 0.8  
(Hummer \& Storey 1987; Storey \& Hummer 1995). 
Lower values of D34 and higher values of D54 imply higher optical depths and densities in the disk.

The 34 low-resolution LAMOST spectra covered all outburst periods of EPIC202060631, 
providing an optimal dataset for calculating D34 and D54. 
During the LD and followed new-QS periods, the D34 values are all larger than 2.7, 
implying optically thin disk. 
The D54 values also indicate an optically thin disk during these phases. 
However, for LR period, although the D34 values generally fall in the optically thick region (D34 $<$ 2.7), 
they exhibit substantial variability. 
Notably, in Fig.~\ref{D34D54}, at least three distinct turning points are observed, as marked by black arrows. 
This behavior suggests multiple episodes of mass injection for the central star into the circumstellar disk. 
This interpretation is consistent with the HDUST model comparision results (see Section 6.1), 
where the disk density undergoes multiple oscillations during the LR period. 
Crucially, the Balmer decrement measurements provide independent supporting evidence, complementing the model predictions.

\subsection{Metallic emission of O\,{\sc i} and Fe\,{\sc ii}}

\begin{figure}
\includegraphics[width=1.0\linewidth]{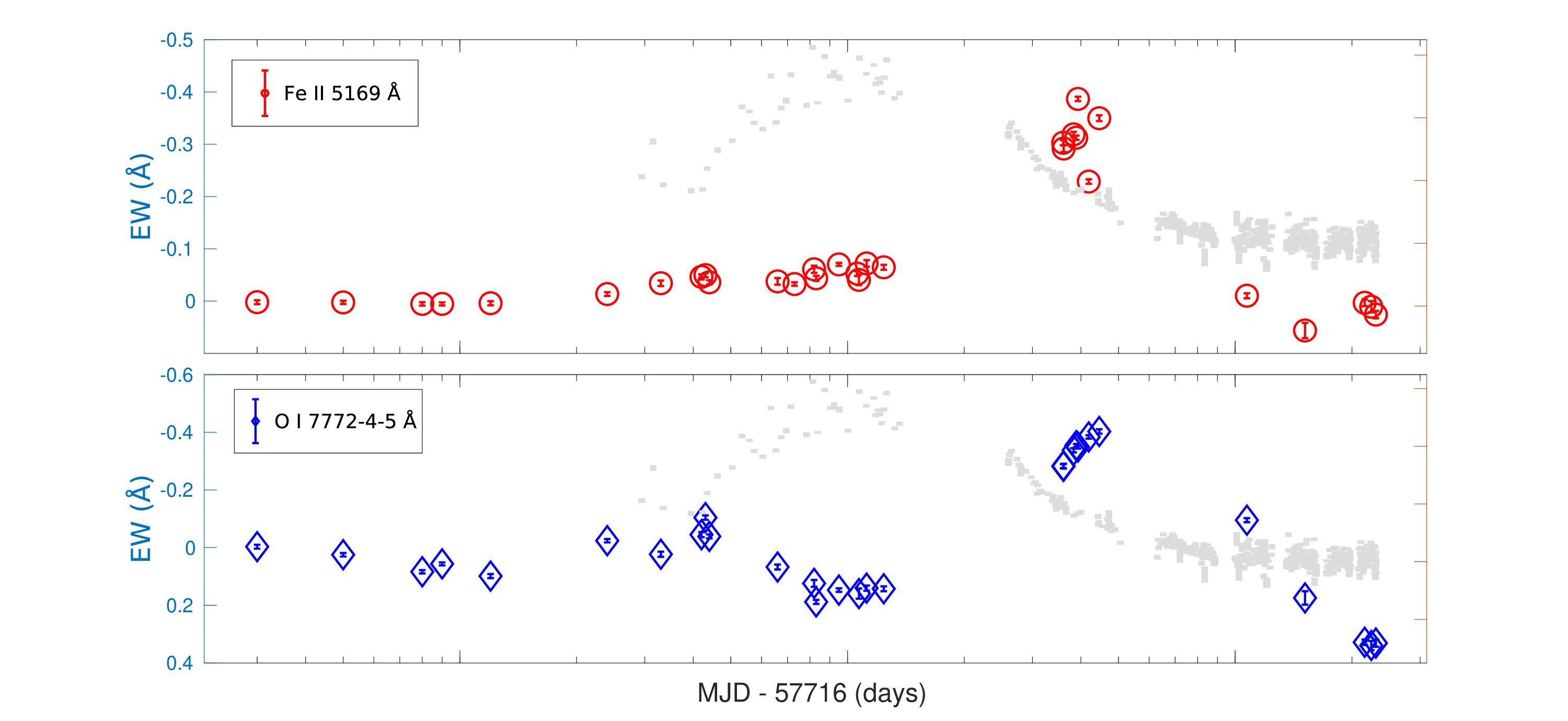}
\caption{\label{MetaEmi} Variations in the equivalent widths (EWs) of the Fe\,{\sc ii} 5169 \AA line (red circles) and the O\,{\sc i} 7772-4-5 \AA 
triplet emissions (blue diamonds) measured from low-resolution spectra. 
Error bars are smaller than the marker sizes and are plotted at the centers. 
The light curve is shown in gray for comparison. 
All values are derived from the low-resolution spectra.
}
\end{figure}

Metallic emission lines were also observed during the outburst of EPIC 202060631. 
The Fe\,{\sc ii} emissions at 5169 and 5317 \AA were visible in both low- and medium-resolution spectra. 
The medium-resolution spectra further revealed Fe\,{\sc ii} emissions at 5018, 5198, 5235, 5276, and 5317 \AA. 

The Fe\,{\sc ii} emission centered at 5169 \AA was selected to calculate the variations in its EWs. 
During the QS period before the outburst, neither emission nor absorption features were present 
in the corresponding spectral region of the 1st low-resolution LAMOST spectrum. 
However, a double-peaked emission feature began to appear in the 10th spectrum during the LR phase, as shown in the top panel of Fig.~\ref{MetaEmi}. The EWs increased monotonically during the LR period and reached maximum levels during the LD phase, exhibiting noticeable dispersion. 
In the final four observations, the emission features disappeared, and the spectral shape reverted to its pre-outburst QS appearance.

Oxygen emission lines were also observed, including the O\,{\sc i} triplet at 7772-4-5 \AA and the 8446 \AA line, 
which is blended with the Paschen P18 emission. 
The EW variations of the O\,{\sc i} line at 7772-4-5 \AA were calculated and are shown in the bottom panel of Fig.~\ref{MetaEmi}. 
Initially, an absorption feature with an EW of 0.37 $\pm$ 0.01 \AA was present. 
Unlike the Fe\,{\sc ii} emissions, the Oxygen line exhibited a decreasing trend in the latter half of the LR period. 
The differing behavior between those emission lines may suggest the circumstellar disk has a more significant impact on the continuum at longer wavelengths around 7000 \AA, compared to shorter wavelengths around 5000 \AA.

The O\,{\sc i} 7772-4-5 triplet is thought to be an indicator of the inner part of the disk, 
with an outer radius of approximately two times the central star's radius. 
Moreover, collisional excitation is regarded as the contributor to this emission 
(Banerjee et al. 2021). 
The spectral line exhibited a narrower emission width during the LD phase but reverted to a complete absorption feature during the new-QS period. 
This may imply that the inner disk continued to expand or fell onto the stellar surface, leaving a relatively vacant space in the inner region.

\section{Discussion and Conclusions}

This work conducted detailed photometric and spectroscopic monitoring of the Be star EPIC 202060631 through its quiescent state, major outburst phase, and return to a new quiescent state. Our main findings include a $\sim$30\% increase in brightness, significant changes in pulsation behavior and amplitudes, indications of radial and tangential acceleration of the photosphere, and clear evidence of radiative heating and mass ejection. These phenomena provide rich clues for investigating the physical origins of Be star outbursts.

We attribute the Be star outburst to a dramatic enhancement of non-radial pulsations, possibly triggered by small perturbations that are amplified in the chaotic regime of the rapidly rotating star. The oblate shape caused by rapid rotation causes material to be preferentially distributed towards the stellar surface, favoring mass ejection. We conjecture that in the early stages of the outburst, violent pulsation enhancements led to multiple small-scale mass ejection events, which later evolved into a large-scale outflow and the formation of a circumstellar disk.

Be stars provide an excellent natural laboratory for studying the effects of rapid rotation and oblate shapes on stellar interiors and dynamics. The various phenomena discovered in this work, such as radiative heating, $v$sin\,$i$ variations, and radial expansion of the photosphere, can serve as observational constraints for testing existing models and developing new ones. Furthermore, we speculate that the mass ejection process in Be stars may be connected to theoretical models of planet formation, as both involve outward material flow and disk formation around a central star, albeit on vastly different scales and timescales.

In summary, this work has revealed the complex and variable nature of Be star outburst activity, while also providing valuable clues for a deeper understanding of this phenomenon. Future continuous observations and numerical simulations, incorporating multi-wavelength data, are needed to further dissect the triggering mechanisms of Be star outbursts and their implications for stellar structure and circumstellar disk formation. Additionally, exploring whether a universal model exists overall the entire evolutionary process of the whole star even into the pulsars stages, such as pulsars, are important topics worthy of investigation.





\bibliographystyle{plainnat}









\textbf{REFERENCE}

Bailer-jones C. A.L., Rybizki.,Fouesneau M., Demleitner M., Andrae R.,
2021, VizieR Online Data Catalog, p. I/352

Banerjee G., Mathew B., Paul K. T., Subramaniam A., Bhattacharyya S.,
Anusha R., 2021, MNRAS, 500, 3926

Carciofi A. C., Bjorkman J. E., 2006, ApJ, 639, 1081

Carciofi A. C., Bjorkman J. E., 2008, ApJ, 684, 1374

Green G. M., Schlafly E. F., Zucker C., Speagle J. S., Finkbeiner D. P., 2019,arXiv e-prints

Rivinius T., Carciofi A. C., Martayan C., 2013, A\&ARv, 21, 69

Vieira R. G., Carciofi A. C., Bjorkman J. E., Rivinius T., Baade D., Rímulo L. R., 2017, MNRAS, 464, 3071


\label{lastpage}
\end{document}